\documentclass[pre,nofootinbib,amsmath,amssymb,twocolumn,showpacs]{revtex4}
\usepackage[T1]{fontenc} \usepackage[latin1]{inputenc}
\usepackage{amsmath} \usepackage{graphicx} \usepackage{amssymb}

\begin{document}

\title{Anomalous Polymer Dynamics Is Non-Markovian:  Memory Effects
and The Generalized Langevin Equation Formulation}

\author{Debabrata Panja}

\affiliation{Institute for Theoretical Physics, Universiteit van
Amsterdam, Valckenierstraat 65, 1018 XE Amsterdam, The Netherlands}

\begin{abstract} Any first course on polymer physics teaches that the
dynamics of a tagged monomer of a polymer is anomalously subdiffusive,
i.e., the mean-square displacement of a tagged monomer increases as
$t^\alpha$ for some $\alpha<1$ until the terminal relaxation time
$\tau$ of the polymer. Beyond time $\tau$ the motion of the tagged
monomer becomes diffusive. Classical examples of anomalous dynamics in
polymer physics are single polymeric systems, such as phantom Rouse,
self-avoiding Rouse, self-avoiding Zimm, reptation, translocation
through a narrow pore in a membrane, and many-polymeric systems such
as polymer melts. In this pedagogical paper I report that all these
instances of anomalous dynamics in polymeric systems are robustly
characterized by power-law memory kernels within a {\it unified\/}
Generalized Langevin Equation (GLE) scheme, and therefore, are
non-Markovian. The exponents of the power-law memory kernels are
related to the relaxation response of the polymers to local strains,
and are derived from the equilibrium statistical physics of
polymers. The anomalous dynamics of a tagged monomer of a polymer in
these systems is then reproduced from the power-law memory kernels of
the GLE via the fluctuation-dissipation theorem (FDT). Using this GLE
formulation I further show that the characteristics of the drifts
caused by a (weak) applied field on these polymeric systems are also
obtained from the corresponding memory kernels.
\end{abstract}


\maketitle

\section{Introduction\label{sec1}} 

One of the most common phrases of wisdom in polymer physics is that a
polymer, in its terminal relaxation time $\tau$, displaces itself by
its own size in physical space \cite{degennes,de}. Being constructed
from monomers connected in series, a polymer has a wide range of
length-scale dependent relaxation times. The longest one of them is
the terminal relaxation time $\tau$, manifested in the decay of the
polymer's end-to-end vector correlation function. [E.g., for a phantom
Rouse polymer of length $N$, the relaxation time corresponding to
length scale $N/p$ scales $\sim (N/p)^2$ with $p$ a positive integer,
implying that $\tau\sim N^2$ for a phantom Rouse polymer]
\cite{de}. The terminal relaxation time $\tau$ of a polymer scales
with its length as a power-law $\tau\sim N^\kappa$ for some exponent
$\kappa$, while its own size in space scales as $N^\nu$ for some
exponent $\nu$. If the polymer is to displace itself by its own size
in time $\tau$, in simplest of cases the mean-square displacement of a
tagged monomer of a polymer must behave $\sim t^{2\nu/\kappa}$, and
since the quantity $2\nu/\kappa$ is not necessarily unity, the dynamics
of a tagged monomer in a polymer must be anomalous till the terminal
relaxation time. If asked to provide canonical examples of such
anomalous behavior, a polymer physicist would almost certainly cite
single polymer dynamics, such as phantom Rouse ($\nu=1/2,\kappa=2$),
self-avoiding Rouse [$\nu\approx0.588$ in three dimensions (3D),
$\nu=3/4$ in two dimensions (2D) and $\kappa=1+2\nu$] and
self-avoiding Zimm ($\nu\approx0.588$ in 3D, $=3/4$ in 2D and
$\kappa=3\nu$) polymers. (The correct single polymer dynamics in a
fluid was first presented by Zimm \cite{zimm}. In this paper I refer
to such polymers, for which the monomers interact with each other via
hydrodynamic interactions, as Zimm polymers. Few years earlier than
Zimm, Rouse \cite{rouse} put forward a model for single polymer
dynamics that neglect the hydrodynamic interactions between the
monomers. Although incorrect, the corresponding (Rouse) polymer
dynamics, for polymers that are self-avoiding, as well as the phantom
ones, i.e., polymers that can intersect themselves, resides at the
heart of polymer physics \cite{de} --- and widely used till today ---
largely due to its simplicity.) Given such an abundance of anomalous
dynamics in polymeric systems, a natural question is whether it is
possible to provide a generic stochastic foundation for it.
\begin{table*}
\begin{tabular}{c|c|c|c} 
  \hline\hline $\quad$polymeric system$\quad$ &
  $\quad$scaling of $\tau\quad$ & $\quad$mean square displacement$\quad$
  & $\quad\mu(t)\quad$\cr \hline\hline phantom Rouse& $\sim N^2$   &
  $\sim t^{1/2}$ till $\tau$ and $\sim t$ thereafter   & $\sim
  t^{-1/2}\exp(-t/\tau)$ \cr\hline self-avoiding Rouse  & $\sim
  N^{1+2\nu}$  & $\sim t^{2\nu/(1+2\nu)}$ till $\tau$ and $\sim t$
  thereafter  & $\sim t^{-2\nu/(1+2\nu)}\exp(-t/\tau)$ \cr\hline
  self-avoiding Zimm  & $\sim N^{3\nu}$  & $\sim t^{2/3}$ till $\tau$
  and $\sim t$ thereafter  & $\sim t^{-2/3}\exp(-t/\tau)$ \cr\hline
  phantom Rouse  & $\sim N^2$ & $\sim t$  & $\sim t^{-1}\exp(-t/\tau)$
  \cr polymer translocation  & ($\tau$ not translocation time)   &   &
  translocation time $\sim N^2$ \cr\hline self-avoiding Rouse  & $\sim
  N^{1+2\nu}$ & $\sim t^{(1+\nu)/(1+2\nu)}$ till $\tau$ & $\sim
  t^{-(1+\nu)/(1+2\nu)}\exp(-t/\tau)$ \cr polymer translocation  &
  ($\tau$ not translocation time)  & and $\sim t$ thereafter  &
  translocation time $\sim N^{2+\nu}$ \cr\hline self-avoiding Zimm  &
  $\sim N^{3\nu}$ & $\sim t^{(1+\nu)/(3\nu)}$ till $\tau$ & $\sim
  t^{-(1+\nu)/(3\nu)}\exp(-t/\tau)$ \cr polymer translocation  & ($\tau$
  not translocation time)  & and $\sim t$ thereafter  & translocation
  time $\sim N^{1+2\nu}$ \cr\hline reptation; repton model & $\sim N^2$
  &$\sim t^{1/2}$ till $\tau$ and $\sim t$ thereafter & $\sim
  t^{-1/2}\exp(-t/\tau)$ \cr (curvilinear co-ordinate)&&&\cr\hline
  &&$\sim t^{1/4}$ between $\tau_e$ and $\sim N^2$ (*) &$\sim t^{-1/4}$
  between $\tau_e$ and $\sim N^2$\cr melt (reptation theory) & $\sim
  N^3$ & $\sim t^{1/2}$ between $\sim N^2$ and $\sim \tau$ & (as much as
  data can resolve)\cr &&and $\sim t$ thereafter&\cr\hline\hline
\end{tabular}
\caption{Summary list of the polymeric systems (polymers of length
  $N$) with anomalous dynamics, and the corresponding memory kernel
  $\mu(t)$, for which the GLE formulation is reported in this paper. (*)
  $\tau_e$ is the so-called entanglement time for polymer melts in the
  reptation theory \cite{de}. Note that if $\mu(t)\sim t^{-\alpha}$,
  then the anomalous dynamics exponent is also $\alpha$.\label{table1}}
\end{table*}

In a recent Letter \cite{panja1}, hereafter referred to as Ref. I, I
reported that the anomalous dynamical behavior for phantom Rouse,
self-avoiding Rouse and Zimm polymers, and that of polymer
translocation through a narrow pore in a membrane can be theoretically
formulated via the following Generalized Langevin Equation (GLE),
wherein the velocity $\vec v(t)$ of a tagged monomer and the force
$\vec\phi(t)$ it experiences, are related to each other via
\begin{eqnarray}   
  \vec\phi(t)=-\int_0^tdt'\mu(t-t')\vec v(t')+\vec
  g(t).
\label{e1}
\end{eqnarray}   
In Eq. (\ref{e1}) $\mu(t)$ is the memory kernel, and the stochastic
noise term $\vec g(t)$ satisfies the condition that $\langle\vec
g(t)\rangle_0=0$, while the fluctuation-dissipation theorem (FDT)
$\langle\vec g(t)\cdot\vec
g(t')\rangle_0\equiv\langle\vec\phi(t)\cdot\vec\phi(t')\rangle_{\vec
  v=0}=3k_BT\mu(t-t')$ in 3D. Here $k_B$ is the Boltzmann constant,
$T$ is the temperature, and $\langle\ldots\rangle_0$ denotes an
average over the stochastic noise realizations, including an average
over equilibrium configurations of the polymers at $t=0$. Equation
(\ref{e1}) can be inverted to write
\begin{eqnarray}   
  \vec v(t)=-\int_0^tdt'a(t-t')\vec\phi(t')+\vec
  h(t),
\label{e2}
\end{eqnarray}    
where $\tilde\mu(s)\tilde a(s)=1$ in the Laplace space, $\langle\vec
h(t)\rangle_0=0$, with the corresponding FDT $\langle\vec
h(t)\cdot\vec h(t')\rangle_0\equiv\langle\vec v(t)\cdot\vec
v(t')\rangle|_{\vec \phi=0}=3k_BT\,a(t-t')$. In Ref. I, I argued that
on the one hand $\mu(t)$ is the mean relaxation response of the
polymers to local strains, and can be derived from the equilibrium
statistical physics of polymers; and on the other, $a(t)$
characterizes the anomalous dynamics via the FDT: as the mean-square
displacement of a tagged monomer is obtained by integrating
$\langle\vec v(t)\cdot\vec v(t')\rangle_{\vec\phi=0}$ twice in
time. An important property of the anomalous dynamics that transpires
through this exercise is that if $\mu(t)\sim t^{-\alpha}$ for some
$\alpha$, then the anomalous dynamics is also $\alpha$. In other
words, according to Ref. I, the anomalous dynamics for phantom Rouse,
self-avoiding Rouse and Zimm polymers, and that of polymer
translocation through a narrow pore in a membrane are connected to the
mean relaxation response of the polymers to local strains.

The purpose of this paper is to provide a pedagogical account of the
unified GLE foundation (\ref{e1}-\ref{e2}) for the classical examples
of anomalous polymer dynamics, including the ones considered in
Ref. I. The precise issues covered in this paper are the following.
\begin{itemize}
\item[(i)] I provide an elaborate derivation of
  Eq. (\ref{e1}-\ref{e2}) for phantom Rouse polymers, which is the
  only fully analytically tractable polymer dynamical system.
\item[(ii)] I substantially supplement the scaling results of Ref. I
  with simulation data, and extend the GLE formulation to the cases of
  a single reptating polymer, and polymer melts.
\item[(iii)] Given the GLE formulation I show that the characteristics
  of the drifts caused by a (weak) applied field on these polymeric
  systems, too, are obtained from the corresponding memory kernels: if
  $\mu(t)\sim t^{-\alpha}$ for some $\alpha$, then the anomalous
  dynamics is also $\alpha$. This could be thought of as the analog of
  the Nernst-Einstein relation \cite{vankampen}.
\item[(iv)] Since the GLE is a description of the trajectories in
  phase space, this paper explicitly brings to light the non-Markovian
  character of the anomalous dynamics for polymeric systems. A natural
  question that then arises is ``Is it possible to provide a
  probabilistic description of these trajectories in phase space?''
  While in this paper I leave this question for further research, I
  note that in limited context --- for the anomalous dynamics of
  polymer translocation through a narrow pore in a membrane ---
  fractional Fokker-Planck equation (fFPE) has been put forward in
  recent times \cite{fFPtrans}. In this equation, an extension of the
  standard Fokker-Planck equation, anomalous dynamics is a consequence
  of introducing power-law waiting times before each jump of the
  concerned particle, as the jump length and the waiting times for any
  jump is obtained from fixed probability distributions, independently
  of their values at previous jumps. It is worth emphasizing (as it
  emerges from the GLE formulation elaborated in this paper) that
  there is no power-law waiting time for the movements of the
  concerned monomer; instead if it makes a move at any time, there is
  an enhanced chance to undo this move in subsequent times; this is
  where anomalous dynamics in polymeric systems stem from. This
  establishes that the fFPE is unsuitable to describe anomalous
  polymer dynamics (polymer translocation included) that are
  considered in this paper.
\end{itemize}

The paper is organized in five sections. Section \ref{sec2} is devoted
to single polymeric systems. Therein I provide (a) an elaborate
derivation of Eq. (\ref{e1}-\ref{e2}) for phantom Rouse polymers, (b)
the characterization of $\mu(t)$ as the mean relaxation response of
the polymers to local strains, and (c) the GLE formulation for
self-avoiding Rouse, self-avoiding Zimm, polymer reptation, and
polymer translocation through a narrow pore in a membrane. In
Sec. \ref{sec3} I provide the GLE formulation for the anomalous
dynamics of a polymer melt, a many-polymeric system. In
Sec. \ref{sec4} I show that the characteristics of the drifts caused
by a (weak) applied field on these polymeric systems are obtained from
the corresponding memory kernels. This could be thought of as an
analogue of the Nernst-Einstein relation. Section \ref{sec5} is
devoted to a critique of recent attempts to model by the fFPE the
anomalous dynamics of polymer translocation through a narrow pore in a
membrane. The paper is finally concluded in Sec. \ref{sec6}.

But before I begin, in Table \ref{table1} I provide a summary list of
the polymeric systems with anomalous dynamics and the corresponding
memory kernel $\mu(t)$ for which the GLE formulation is argued in this
paper.

\section{The GLE formulation for anomalous dynamics in single
polymeric systems\label{sec2}}

\subsection{The GLE formulation for phantom Rouse
polymers\label{sec2a}}

\subsubsection{Derivation of Eq. (\ref{e1}) from the Rouse
equation\label{sec2a1}}

Consider a phantom Rouse polymer of length $N$. It is described by
the Rouse equation; in continuum representation it reads \cite{de}
\begin{eqnarray}      
  \gamma\frac{\partial\vec r_n(t)}{\partial t}=k\,
  \frac{\partial^2 \vec r_n(t)}{\partial n^2}+\vec f_n(t),
\label{e3}
\end{eqnarray}      
where $\vec r_n(t)$ is the location of the $n$-th monomer at time $t$,
$\gamma$ is the damping coefficient of the surrounding fluid, and $k$
is the spring constant for the springs connecting the consecutive
monomers. The stochastic force $\vec f_m(t)$ satisfies the conditions
$\langle\vec f_n(t)\rangle=0$ and the FDT $\langle
f_{m\sigma}(t)f_{n\lambda}(t')\rangle=2\gamma
k_BT\delta(m-n)\delta_{\sigma\lambda}\delta(t-t')$, for
$\sigma,\lambda=(x,y,z)$. Equation (\ref{e3}) is supplemented by the
``open'' boundary conditions that the chain tension of the polymer at
the free ends must vanish; i.e., $(\partial\vec r_n/\partial
n)|_{n=0}=(\partial\vec r_n/\partial n)|_{n=N}=0$.

Since the Rouse equation is linear in $\vec r_n(t)$, it can be solved
to obtain all correlation functions using the mode expansion technique
\cite{de}. Two noteworthy results borne out of this exercise are: (a)
the terminal relaxation time $\tau=\gamma N^2/(\pi^2k)$, and (b) the
MSD of the middle monomer increases as $t^{1/2}$ until time $\tau$,
and only after that time the motion of the middle monomer becomes
diffusive, with the diffusion coefficient scaling $\sim1/N$. It is
this subdiffusive motion of the middle monomer that I obtain from the
GLE (\ref{e1}-\ref{e2}), which in turn I derive exactly from
Eq. (\ref{e3}). For this problem $\vec\phi(t)$ given by the net force
it experiences (forces exerted on the middle monomer by the ones it is
connected to, or equivalently, the chain tensions at the middle
monomer), i.e.,
\begin{eqnarray}   
  \vec\phi(t)=k\left[\frac{\partial\vec
      r_n(t)}{\partial n}\bigg|_{n=(N/2)+}-\frac{\partial\vec
      r_n(t)}{\partial n}\bigg|_{n=(N/2)-}\right].
\label{e4}
\end{eqnarray}   Note that using Eq. (\ref{e4}) the Rouse equation for
the middle monomer can be written as
\begin{eqnarray}        \gamma\vec v(t)=\vec\phi(t)+\vec f_{N/2}(t).
\label{e4a}
\end{eqnarray}     

Given Eqs. (\ref{e1}-\ref{e4a}), an issue that may come to mind is the
following. While the GLE formulation (\ref{e1}-\ref{e2}) has a
distinct memory kernel --- I have shown in Ref. I that $\mu(t)\sim
t^{-1/2}\exp(-t/\tau)$, and similarly $a(t)\sim t^{-3/2}\exp(-t/\tau)$
for phantom Rouse polymers --- Eqs. (\ref{e3}-\ref{e4a}) do not: e.g.,
in Eq. (\ref{e4a}) the velocity of the middle monomer at any time is
simply proportional to the forces at the same time exerted by the
monomers it is connected to (plus a $\delta$-correlated random
noise). From this it may appear that the GLE formulation
(\ref{e1}-\ref{e2}) is in contradiction with
Eqs. (\ref{e3}-\ref{e4a}). In truth however, there is no contradiction
among Eqs. (\ref{e1}-\ref{e4a}). First of all, the fact that there is
no explicit memory kernel in the Rouse equation does not mean that a
phantom polymer does not have memory. As noted in the very first
paragraph of the introduction, the relaxation time corresponding to a
length scale $N/p$ scales $\sim(N/p)^2$ with $p$ an integer \cite{de}
shows that a phantom polymer {\it does\/} have memory; in fact, below
I will show that the memory kernel $\mu(t)$ is indeed built from the
long relaxation times for fluctuations at long length scales of the
polymer. [In this context, it is worth recalling the work by Zwanzig
\cite{zwanzig}, wherein the memory of the concerned particle arises
due to its coupling to a thermal bath of harmonic oscillators,
effected via a Hamiltonian. For the present case, the fluctuation
modes effectively play the role of a thermal bath that is coupled to
the motion of a monomer due to chain connectivity, giving rise to the
memory kernel $\mu(t)$.]  Secondly, while Eqs. (\ref{e1}) and
(\ref{e2}), being inverse Laplace transform of each other, describe
exactly one relation between $\vec v(t)$ and $\vec\phi(t)$,
Eq. (\ref{e4a}) provides the second relation between the two
quantities, so that out of Eqs. (\ref{e1}-\ref{e2}) and (\ref{e4a})
one can form closed equations for both $\vec v(t)$ and
$\vec\phi(t)$. Indeed, in Sec. \ref{sec2a3} I use the closed form
equation for $\vec\phi(t)$ to obtain the velocity autocorrelation
function (of course, it is the same velocity autocorrelation function
as obtained in Ref. I, and as summarized in the introduction of this
paper).

Without further ado, below I derive provide the GLE (\ref{e1}) for
phantom polymers from the Rouse equation (\ref{e3}).

Three ingredients are necessary to derive Eq. (\ref{e1}). The first
one of them is the dynamics for both halves of the polymer when the
middle monomer held fixed at, say, $\vec {\cal R}$. Being a phantom
polymer, when the middle monomer held fixed, the two halves of the
polymer evolve independently of each other. With $\vec r\,'_n(t)=\vec
r_n(t)-\vec{\cal R}$, the corresponding equations for each half of the
phantom polymer satisfy the (Rouse) equation
\begin{eqnarray}      \gamma\frac{\partial\vec r \,'_n}{\partial
t}=k\, \frac{\partial^2 \vec r \,'_n(t)}{\partial n^2}+\vec f_n(t),
\label{e5}
\end{eqnarray}       for $n\neq0$. Equation (\ref{e5}) is supplemented
by the ``open'' boundary conditions that the chain tension of the
phantom polymer at the free ends must vanish; i.e., $(\partial\vec
r_n/\partial n)|_{n=0}=(\partial\vec r_n/\partial n)|_{n=N}=0$
(additionally $\vec r \,'_0=0$ by definition). In order to analyze the
motion of the two halves of the phantom polymer I define
\begin{eqnarray}      \vec Y^{(r)}_p(t)=\frac1N\int_{0}^{\frac
N2}dn\,\sin\frac{\pi (2p+1)n}N\, \vec r\,'_{n+N/2}(t),
\label{e6}
\end{eqnarray} 
\begin{eqnarray}     \mbox{and}\,\,\vec
Y^{(l)}_p(t)=-\frac1N\int_{-\frac N2}^0 dn\, \sin\frac{\pi (2p+1)n}N\,
\vec r\,'_{n+N/2}(t)
\label{e7}
\end{eqnarray}   for $p=0,1,\ldots$, for the right and the left half,
such that
\begin{eqnarray}    \vec r\,'_{n+N/2}(t)=4\sum_p\vec
Y^{(r)}_p(t)\sin\frac{\pi(2p+1)n}N\,\Theta(n)\nonumber\\&&\hspace{-5.6cm}-4\sum_p\vec
Y^{(l)}_p(t)\sin\frac{\pi(2p+1)n}N\,\Theta(-n).
\label{e8}
\end{eqnarray}     Note that Eq. (\ref{e8}) is consistent with the
open boundary conditions, as well as the fact that the middle monomer
remains fixed. The independent evolution of each half satisfies the
Langevin equation (LE)
\begin{eqnarray}   \gamma_p\frac{\partial\vec Y_p}{\partial
t}=-k_p\,\vec Y_p(t)+\vec f_p(t),
\label{e9}
\end{eqnarray}   for $\vec Y_p=[\vec Y^{(l)}_p,\vec Y^{(r)}_p]$,
$\gamma_p=2N\gamma$, $k_p=2\pi^2k(2p+1)^2/N$, $\langle\vec
f_p\rangle=0$ and $\langle
f_{p\sigma}(t)f_{q\lambda}(t')\rangle=\gamma_pk_BT\delta(t-t')\delta_{pq}\delta_{\sigma\lambda}$.

The second ingredient is the equilibrium averages of $Y_p$  and
$Y_p^2$. First, $\langle\vec Y^{(r)}_p(t)\rangle=\langle \vec
Y^{(l)}_p(t)\rangle\equiv0$ by isotropy. Secondly, by left-right
symmetry
$\langle[Y^{(r)}_p(t)]^2\rangle=\langle[Y^{(l)}_p(t)]^2\rangle$;  and
they are obtained from the LE (\ref{e9}) as
\begin{eqnarray}
\hspace{-2mm}\langle[Y^{(r)}_p(t)]^2\rangle=\langle[Y^{(l)}_p(t)]^2\rangle=
3Nk_BT/[4\pi^2k(2p+1)^2].
\label{e10}
\end{eqnarray}

The third ingredient is that in terms of the mode amplitudes $\vec
Y^{(l)}_p(t)$ and $\vec Y^{(r)}_p(t)$ with the middle monomer fixed at
$\vec{\cal R}$, the force $\vec\phi(t)$ on the middle monomer in
Eq. (\ref{e4}) is given by
\begin{eqnarray}  \vec\phi(t)=4k\sum_p\frac{\pi(2p+1)}N\left[\vec
Y^{(r)}_p(t)+\vec Y^{(l)}_p(t)\right].
\label{e11}
\end{eqnarray} 

With these ingredients, I consider an ensemble of phantom polymers
(let us label it as ensemble ${\mathbf X}$ for future reference) in
equilibrium at time $0^-$, such that for each of the polymers in this
ensemble the velocity history of the middle monomer up to time $t>t_m$
is given by
\begin{eqnarray}  \vec v(t>t_m)=\sum_{i=0}^{m}\vec{\delta
r}_i\,\delta(t-t_i)
\label{e12}
\end{eqnarray}  for some $m$ and a sequence $\{\vec{\delta r_i}\}$,
with $t_0=0$. [One does not need not be alarmed that in
Eq. (\ref{e12}) the velocity is represented by a sum over
$\delta$-functions in time: such a representation only simplifies the
following analysis.] Note here, using Eq. (\ref{e6}), (\ref{e7}) and
(\ref{e12},) that
\begin{eqnarray}  \vec Y_p(t_i^+)=\vec Y_p(t_i^-)-\frac2N\,\vec{\delta
r_i}\quad\mbox{for}\,\,i=0,\ldots,m.
\label{e13}
\end{eqnarray} 

Let us now determine the evolution of the forces the middle monomer of
a phantom polymer, belonging to the ensemble ${\mathbf X}$,
experiences as a function of time. With the velocity history as in
Eq. (\ref{e12}), using Eq. (\ref{e9}), the mode amplitudes for a
member polymer of the ensemble evolve as
\begin{eqnarray}   \vec Y_p(t>t_m)\!=\!e^{-k_p(t-t_m)/\gamma_p}\vec
Y_p(t_m^+)\!\nonumber\\&&\hspace{-3cm}+\!\frac1\gamma_p\!\int_{t_m}^t
\!\!dt'\, e^{-k_p(t-t')/\gamma_p}f_p(t')
\label{e14}
\end{eqnarray}  for $\vec Y_p=[\vec Y^{(l)}_p,\vec Y^{(r)}_p]$.
Thereafter, using Eq. (\ref{e11}-\ref{e14})
\begin{eqnarray} \frac{\vec\phi(t>t_m)}k\!=\!-\frac8N\vec{\delta
r}_m\!\sum_p\!e^{-k_p(t-t_m)/\gamma_p}\nonumber\\
&&\hspace{-4cm}+4\sum_p\!\frac{\pi(2p+1)}N\vec g^{\,(m)}_p(t),\,
\mbox{with}\nonumber\\ &&\hspace{-6.5cm}\vec
g^{\,(m)}_p(t)=\!e^{-k_p(t-t_m)/\gamma_p}[\vec Y^{(r)}_p(t_m^-)+\vec
Y^{(l)}_p(t_m^-)]\nonumber\\
&&\hspace{-5.5cm}+\frac1\gamma_p\int_{t_m}^t \!\!dt'\,
e^{-k_p(t-t')/\gamma_p}[f^{(r)}_p(t')+f^{(l)}_p(t')].
\label{e15}
\end{eqnarray}  Further, as I extend Eqs. (\ref{e13}-\ref{e14})
further back in time to write
\begin{eqnarray} [\vec Y^{(r)}_p(t_m^-)+\vec
Y^{(l)}_p(t_m^-)]=-\frac2N\,\vec{\delta
r}_{m-1}\,e^{-k_p(t_m-t_{m-1})/\gamma_p}\nonumber\\
&&\hspace{-7.7cm}+\,e^{-k_p(t_m-t_{m-1})/\gamma_p}[\vec
Y^{(r)}_p(t_{m-1}^-)+\vec Y^{(l)}_p(t_{m-1}^-)]\nonumber\\
&&\hspace{-7.7cm}+\,\frac1\gamma_p\,\int_{t_{m-1}}^{t_m} \!\!dt'\,
e^{-k_p(t_m-t')/\gamma_p}[f^{(r)}_p(t')+f^{(l)}_p(t')],
\label{e16}
\end{eqnarray}  Eq. (\ref{e15}) becomes
\begin{eqnarray} \frac{\vec\phi(t>t_m)}k\!=\!-\frac8N\vec{\delta
r}_m\!\sum_p\!e^{-k_p(t-t_m)/\gamma_p}\nonumber\\
&&\hspace{-4cm}-\frac8N\vec{\delta
r}_{m-1}\!\sum_p\!e^{-k_p(t-t_{m-1})/\gamma_p}\nonumber\\
&&\hspace{-4cm}+4\sum_p\!\frac{\pi(2p+1)}N\vec g^{\,(m-1)}_p(t),\,
\mbox{with}\nonumber\\ &&\hspace{-6.1cm}\vec
g^{\,(m-1)}_p(t)=\!e^{-k_p(t-t_{m-1})/\gamma_p}[\vec
Y^{(r)}_p(t_{m-1}^-)+\vec Y^{(l)}_p(t_{m-1}^-)]\nonumber\\
&&\hspace{-5.5cm}+\frac1\gamma_p\int_{t_{m-1}}^t \!\!dt'\,
e^{-k_p(t-t')/\gamma_p}[f^{(r)}_p(t')+f^{(l)}_p(t')].
\label{e17}
\end{eqnarray}   At this point, a comparison between Eqs. (\ref{e15})
and (\ref{e17}) shows that the force $\phi(t>t_m)$ on the middle
monomer of a polymer, belonging to ensemble ${\mathbf X}$, can be
extended all the way to time $0^-$, to write
\begin{eqnarray}
\frac{\vec\phi(t>t_m)}k\!=\sum_{i=0}^m\underbrace{\left[-\frac8N\vec{\delta
r}_i\!\sum_p\!e^{-k_p(t-t_i)/\gamma_p}\right]}_{\vec
q_i(t)}\nonumber\\
&&\hspace{-4cm}+\underbrace{4\sum_p\!\frac{\pi(2p+1)}N\vec
g^{\,(0)}_p(t)}_{\vec g(t)},\, \mbox{with}\nonumber\\
&&\hspace{-7cm}\vec g^{\,(0)}_p(t)=\!e^{-k_pt/\gamma_p}[\vec
Y^{(r)}_p(0^-)+\vec Y^{(l)}_p(0^-)]\nonumber\\
&&\hspace{-6cm}+\frac1\gamma_p\int_{0}^t \!\!dt'\,
e^{-k_p(t-t')/\gamma_p}[f^{(r)}_p(t')+f^{(l)}_p(t')].
\label{e18}
\end{eqnarray} 

It is now seen, by converting the sum to an integral, that
\begin{eqnarray}    \vec q_i(t)=-\frac{8k}N\vec{\delta r}_i\sum_p
e^{-k_p(t-t_i)/\gamma_p}\nonumber\\ &&\hspace{-4.5cm}=-2\,\vec{\delta
r}_i\, \sqrt{\frac{\pi\gamma k} {t-t_i}}\, e^{-(t-t_i)/\tau}.
\label{e19}
\end{eqnarray}    It is also seen, using Eq. (\ref{e10}), that
$\langle\vec g(t)\rangle_0=0$, with the FDT
\begin{eqnarray}    \langle\vec g(t)\cdot\vec
g(t')\rangle_0=\frac{24kk_BT}N\sum_p\!e^{-k_pt/\gamma_p}\nonumber\\
&&\hspace{-4cm}=6k_BT\sqrt{\frac{\pi\gamma k}{(t-t')}}\,
e^{-(t-t')/\tau}
\label{e20}
\end{eqnarray}    assuming $t>t'$. Here $\langle\ldots\rangle_0$
denotes an average over the noise realizations, including an average
over the configurations of the polymers of the  ensemble ${\mathbf X}$
at $t=0^-$. The equilibrium condition for ensemble ${\mathbf X}$ at
$t=0^-$ is necessary in Eqs. (\ref{e19}-\ref{e20}) to have
$\langle\vec Y^{(r)}_p(t)\rangle_{0^-}=\langle\vec
Y^{(l)}_p(t)\rangle_{0^-}\equiv0$ by isotropy, and also to provide us
with
$\langle[Y^{(r)}_p(t)]^2\rangle_{0^-}=\langle[Y^{(l)}_p(t)]^2\rangle_{0^-}=3Nk_BT/[4\pi^2k(2p+1)^2]$
[these are the results of Eq. (\ref{e10})]. In other words, by means
of Eqs. (\ref{e18}-\ref{e20}), I have derived the GLE (\ref{e1}) for
the motion of the middle monomer of a phantom Rouse polymer, with
\begin{eqnarray}   \mu(t)=2\sqrt{\frac{\pi\gamma
k}{t}}\,\exp(-t/\tau).
\label{e21}
\end{eqnarray}   
As claimed earlier, note that a comparison of Eq. (\ref{e21}) with
Eq. (\ref{e19}) shows that $\mu(t)$ is indeed built from the polymer's
memory, i.e., from the long relaxation times for fluctuations at long
length-scales. In this context, it is worth recalling the work by
Zwanzig \cite{zwanzig}, wherein the memory of the concerned particle
arises due to its coupling to a thermal bath of harmonic oscillators,
effected via a Hamiltonian. For the present case, the memory kernel
$\mu(t)$ arises due to the physical connectivity of the monomers;
nevertheless, the fluctuation modes of the polymer effectively play
the role of a thermal bath coupled to the motion of a monomer.

Henceforth I drop the subscript `0' from the angular brackets; when
applicable, an average over an equilibrium ensemble of polymers at
$t=0$ will be understood.

\subsubsection{Anomalous dynamics for phantom polymers, and the
derivation of Eq. (\ref{e2}) for phantom polymers \label{sec2a3}}

Having derived Eq. (\ref{e1}) for phantom polymers in
Sec. \ref{sec2a1}, I now proceed to characterize the anomalous
dynamics for the middle monomer of a phantom polymer using
Eq. (\ref{e4a}). Together, Eqs. (\ref{e1}) and (\ref{e4a}) yield
\begin{eqnarray}  
  \gamma\vec v(t)=-\int_0^tdt'\,\mu(t-t')\vec
  v(t')+\underbrace{[\vec g(t)+\vec f_{N/2}(t)]}_{\vec p(t)}.
\label{e22}
\end{eqnarray}   
 I then use Laplace transform to express
\begin{eqnarray}  
  \vec v(t)=\int_0^tdt'\,\beta(t-t')[\vec g(t')+\vec
  f_{N/2}(t')],
\label{e23}
\end{eqnarray}  where in the Laplace space
$\tilde\beta(s)[\gamma+\tilde\mu(s)]=1$. Note that the noise term
$\vec p(t)$ in Eq. (\ref{e22}) also satisfies the FDT, as it should:
the $\vec f_{N/2}(t)$ part accounts for the viscous work involving
$\gamma$ [this follows from the Rouse equation (\ref{e3})], while the
$\vec g(t)$ accounts for $\mu(t)$, as in Eq. (\ref{e21}).

Before proceeding further I make the observation that
Eqs. (\ref{e22}-\ref{e23}) are closed w.r.t. the velocity $\vec v(t)$
of the middle monomer. In fact, Eqs. (\ref{e22}-\ref{e23}) provide the
trajectory description in the $(\vec r, \vec v)$ phase space, and
clearly demonstrates that {\it the stochastic process underlying the
  anomalous dynamics for phantom polymers is non-Markovian}. I will
return to this issue in Sec. \ref{sec4}.

The calculation for $\langle\vec v(t)\cdot\vec v(t')\rangle$ based on
Eq. (\ref{e23}) can be found in the Appendix. I quote the result
below, assuming $t>t'$:
\begin{eqnarray}  
\langle\vec v(t)\cdot\vec
v(t')\rangle=3k_BT\,\beta(t-t'),
\label{e24}
\end{eqnarray}  
where $\beta(t)$ can be obtained by Laplace inverting
the relation $\tilde\beta(s)[\gamma+\tilde\mu(s)]=1$. Since from
Eq. (\ref{e21}), $\tilde\mu(s)=2\sqrt{\pi\gamma
k}(s+\tau^{-1})^{-1/2}$, and therefore
$\tilde\beta(s)=[\gamma+2\sqrt{\pi\gamma
k}(s+\tau^{-1})^{-1/2}]^{-1}$. Consequently, at long times, i.e., for
times $(t-t')\gg\gamma/k$ but $(t-t')\ll\tau$ [this is possible in the
limit of large $N$: recall that $\tau=\gamma N^2/(\pi^2k)$], the
relation $\tilde\beta(s)[\gamma+\tilde\mu(s)]=1$ in the Laplace space
can be inverted to obtain
\begin{eqnarray}  
  \langle\vec v(t)\cdot\vec
  v(t')\rangle\sim(t-t')^{-3/2}e^{-(t-t')/\tau}\!\!.
\label{e25}
\end{eqnarray}  
Subsequently, the result that the mean-square displacement (MSD) of
the middle monomer increases $\sim t^{1/2}$ till time $\tau$ and $\sim
t$ thereafter is obtained by integrating of $\langle\vec v(t)\cdot\vec
v(t')\rangle$ twice in time.

It is important to note here that the following result transpires
through the above exercise: if $\mu(t)\sim t^{-\alpha}$ for some
$\alpha$, then the MSD of the middle monomer will increase as
$t^\alpha$.

Interestingly, instead of going through the
Eqs. (\ref{e22}-\ref{e24}), one can choose an alternative route to
arrive at Eq. (\ref{e25}) that I took in Ref. I. Given the GLE
(\ref{e1}), one arrives at Eq. (\ref{e2}), namely, at
\begin{eqnarray}   \vec v(t)=-\int_0^tdt'a(t-t')\vec\phi(t')+\vec
h(t),
\label{e26}
\end{eqnarray}    with $\tilde\mu(s)\tilde a(s)=1$ in the Laplace
space, and the corresponding FDT $\langle\vec h(t)\cdot\vec
h(t')\rangle_0\equiv\langle\vec v(t)\cdot\vec v(t')\rangle|_{\vec
\phi=0}=3k_BT\,a(t-t')$. Thereafter, in order to obtain the velocity
autocorrelation function (\ref{e25}) from Eq. (\ref{e26}), one
proceeds along the following manner. From the Rouse equation
(\ref{e3}) or (\ref{e4}), one needs to appreciate that $\gamma/k$ is
proportional to the middle monomer's reaction time to balance out the
chain tensions $\displaystyle{\frac{\partial\vec r_n(t)}{\partial
n}\bigg|_{n=(N/2)+}}$ and $\displaystyle{\frac{\partial\vec
r_n(t)}{\partial n}\bigg|_{n=(N/2)-}}$. Thus, for $(t-t')\gg\gamma/k$
one is always at a limit $\phi\approx0$, implying that
\begin{eqnarray}  \langle\vec v(t)\cdot\vec
v(t')\rangle\approx\langle\vec v(t)\cdot\vec
v(t')\rangle_{\vec\phi=0}=3k_BT\,a(t-t'),
\label{e27}
\end{eqnarray}  i.e., I arrive back at Eq. (\ref{e25}) \cite{laplace},
which I have elaborated via Eqs. (\ref{e3}-\ref{e24}).

To summarize Sec. \ref{sec2} so far: (i) up to Eq. (\ref{e25}) I have
provided an elaborate derivation for the GLE formulation for the
anomalous dynamics of phantom polymers, explicitly demonstrating that
the stochastic process underlying the anomalous dynamics of the middle
monomer is non-Markovian. (ii) Given that the GLE formulation
(\ref{e1}-\ref{e2}) dictates that if $\mu(t)\sim t^{-\alpha}$, then
the anomalous dynamics exponent is also $\alpha$, characterization of
$\mu(t)$ is sufficient to lead one to the anomalous dynamical behavior
of phantom polymers.

In the following subsections \ref{sec2b}-\ref{sec2d} and in
Sec. \ref{sec3} I will consider the anomalous dynamics for
self-avoiding Rouse and Zimm polymers, polymer translocation through a
narrow pore in a membrane, polymer reptation and polymer melts. These
systems are not analytically tractable like phantom polymers have
been. Nevertheless, as I have demonstrated above for phantom polymers,
and summarized in Ref. I, characterizing $\mu(t)$ will turn out to be
powerful enough to arrive at the anomalous dynamical behavior for all
these polymeric systems via the GLE description
(\ref{e1}-\ref{e2}). To this end, one needs to identify the generic
properties involving $\mu(t)$ and FDT; these are carried out in
Sec. \ref{sec2a2} below.

\subsubsection{Three generic issues for polymeric systems: (a)
$\mu(t)$ as the polymer's mean relaxation response to local strain,
(b) fluctuation-dissipation theorem, and (c) Eq. (\ref{e22}) for
polymeric systems in general \label{sec2a2}}

I begin by having observed the following from Eq. (\ref{e1}): $\mu(t)$
is the mean response of the polymer to a local strain. [For phantom
polymers, the strain is due to the alteration in the chain tensions
$\displaystyle{k\frac{\partial\vec r_n(t)}{\partial
n}\bigg|_{n=(N/2)+}}$ and $\displaystyle{k\frac{\partial\vec
r_n(t)}{\partial n}\bigg|_{n=(N/2)-}}$, caused by moving the middle
monomer by a distance $\vec{\delta r}$ at time $t_0$ and fixing it at
its new position $\forall t$. Then on average, the local strain then
relaxes in time $\sim (t-t_0)^{-1/2}$, i.e.,
$\langle\vec\phi(t)\cdot\vec\phi(t_0)\rangle_{\vec
v=0}\sim(t-t_0)^{-1/2}$.] Based on this observation, I now discuss
three generic issues that would be instrumental for arriving at the
anomalous dynamics for polymeric systems, as these systems are not
analytically tractable like phantom polymers have been.
\begin{itemize}
\item[(a)] While the mean response of a polymer to a local strain
  depends on how the strain is created, {\it the identification of
    $\mu(t)$ as the polymer's mean local strain relaxation response
    alone allows one to write down the GLE\/}, as I argue
  below. First, given the identification of $\mu(t)$ as the polymer's
  mean local strain relaxation response one can always write the
  stochastic Eq. (\ref{e1}) with $\langle\vec g(t)\rangle=0$, which
  holds by definition. Next, to obtain the FDT, consider
  Eq. (\ref{e1}) for an ensemble of polymers with $\vec
  v(t)=0\,\,\forall t$ and $\vec\phi(t_0)=\vec g_0$, a specific
  value. For such an ensemble $\vec g(t)\equiv\vec\phi(t)$, and since
  $\mu(t)$ is the polymer's mean local strain relaxation response,
  $\langle\vec\phi(t)\cdot\vec\phi(t_0)\rangle=g^2_0\,\mu(t-t_0)$ for
  $t>t_0$. Extending this to the dynamics of a polymer in an
  equilibrium ensemble (where $\vec g_0$ is also chosen from the
  equilibrium ensemble), one has $\langle\vec g(t)\cdot\vec
  g(t_0)\rangle\equiv\langle\vec\phi(t)\cdot\vec\phi(t_0)\rangle_{\vec
    v=0}=\langle \phi^2(t)\rangle_{\vec v=0}\,\mu(t-t_0)$.
\item[(b)] Conversely --- i.e., turning the argument of (a) around ---
  the existence of the FDT $\langle\vec g(t)\cdot\vec
  g(t_0)\rangle=C\,\mu(t-t_0)$ for some constant $C$, along with
  $\langle\vec g(t)\rangle=0$ also implies that the polymer's mean
  relaxation response to local strains --- created at time $t_0$ ---
  is $\propto\mu(t-t_0)$. Let us assume that the polymer's mean
  relaxation response to local strains, created at time $t_0$, is
  given by $\eta(t-t_0)$ for some function $\eta$. Using (a) above,
  one deduces that the corresponding FDT will have to take the form
  $\langle\vec g(t)\cdot\vec g(t_0)\rangle=\langle\vec
  g^2(t_0)\rangle\eta(t-t_0)$, which, by construction equals
  $C\mu(t-t_0)$; i.e., $\eta(t)\propto\mu(t)$.
\item[(c)] In addition, for a tagged monomer in a polymeric system,
  one anticipates Eq. (\ref{e4a}) to generically hold true, with
  $\gamma$ being an effective damping coefficient of the surrounding
  medium (that includes the interactions with other
  monomers!). Together, they then yield Eq. (\ref{e22}), which is a
  closed form equation for $\vec v(t)$. It is Eq. (\ref{e22}) that
  tells us that if the monomer makes a move at any time, there is an
  enhanced chance to undo this move in subsequent times; this is where
  anomalous dynamics in polymeric systems stem from.
\end{itemize} 
In Secs. \ref{sec2b}-\ref{sec2d} and  \ref{sec3}, in
order to arrive at the GLE formulation (\ref{e1}-\ref{e2}) for the
anomalous dynamics for self-avoiding Rouse and Zimm polymers, polymer
translocation through a narrow pore in a membrane, polymer reptation
and polymer melts, I will use (a-c) above, since these systems are not
analytically tractable like phantom polymers have been. In some cases
I will obtain $\mu(t)$ directly, and in others I will calculate
$\langle\vec\phi(t)\cdot\vec\phi(t')\rangle_{\vec v=0}$. Wherever
possible, I will do both.

\subsection{The GLE formulation (\ref{e1}-\ref{e2}) for the anomalous
dynamics of self-avoiding Rouse and Zimm polymers\label{sec2b}}

The monomers of a self-avoiding polymer interact over a long-range,
which prohibits one from writing down an exact equation for the
velocities of the monomers in terms of the forces they
experience. However, some properties of self-avoiding polymers are
well-known; two of them I need here for a polymer of length $N$ are:
(i) the terminal time $\tau$ scales $\sim N^{1+2\nu}$ for a Rouse
polymer, and as $\sim N^{3\nu}$ for a Zimm polymer \cite{de}; and (ii)
the entropic spring constant of a polymer scales as $N^{-2\nu}$
\cite{degennes}. Here $\nu$ is the Flory exponent, in 3D
$\nu\approx0.588$, and in 2D $\nu=3/4$. With these properties I now
characterize $\mu(t)$ as the polymers' mean local strain relaxation
response, as described in Sec. \ref{sec2a2}.
\begin{figure}[h]
\begin{center}
\includegraphics[angle=270,width=\linewidth]{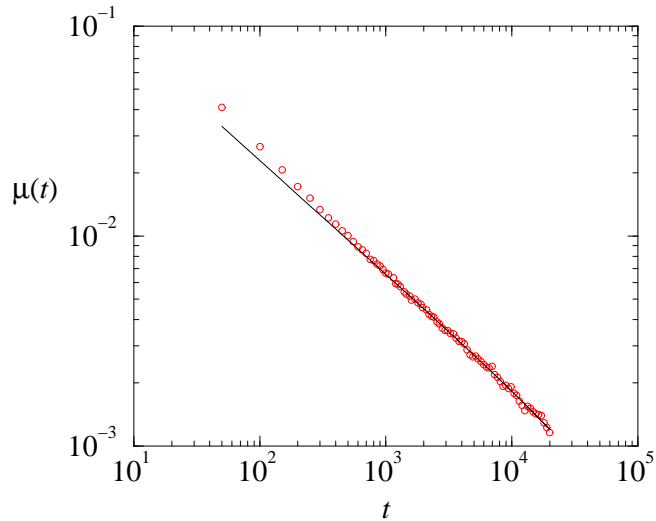}
\end{center}
\caption{(color online) Simulation data for $\mu(t)$ for a
  self-avoiding Rouse polymer in a Monte Carlo lattice polymer model
  for $N=400$, obtained from Eq. (\ref{e22}) and the FDT (see text for
  details). With $2\nu/(1+2\nu)\approx0.54$ in 3D the solid (black)
  curve corresponds to $\mu(t)=0.275t^{-0.54}\exp(-t/\tau)$, where
  $\tau$ was measured to be $\approx223801$. Figure reproduced from
  Ref. \cite{rousepaper} with permission from American Institute of
  Physics.\label{fig1}}
\end{figure}

Imagine that one moves the middle monomer of a self-avoiding polymer
by a small distance $\vec{\delta r}$ at $t=0$ and holds it at its new
position $\forall t>0$, corresponding to $\vec{v}(t)=\vec{\delta
r}\,\delta(t)$. Following the terminal time scaling for self-avoiding
polymers (i), at time $t$, counting away from the middle monomer, all
the monomers within a backbone distance $n_t\sim t^{1/(1+2\nu)}$ for a
Rouse, and $\sim t^{1/(3\nu)}$ for a Zimm polymer equilibrate to the
new position of the middle monomer. However, since the rest $(N-n_t)$
monomers are not equilibrated to the new position of the middle
monomer at time $t$, these $n_t$ monomers are stretched by a distance
$\vec{\delta r}$. With the entropic spring constant of these $n_t$
equilibrated monomers scaling $\sim n_t^{-2\nu}$ [following property
(ii)], the mean force the middle monomer will experience at its new
position is given by $\vec\phi(t)\sim n_t^{-2\nu}(-\vec{\delta r})\sim
t^{-2\nu/(1+2\nu)}(-\vec{\delta r})$ for a Rouse, and $\vec\phi(t)\sim
n_t^{-2\nu}(-\vec{\delta r})\sim t^{-2/3}(-\vec{\delta r})$ for a Zimm
polymer [force $=$ (spring constant) $\times$ (stretching
distance)]. This power-law behavior lasts only till the terminal time
$\tau$. [for a phantom polymer, the time behavior of Eq. (\ref{e21})
is recovered from this of argument upon simply replacing $\nu$ by
$1/2$.] In the light of point (a) in Sec. \ref{sec2a2}, the above
leads one to the GLE formulation (\ref{e1}) for self-avoiding Rouse
and Zimm polymers, with the corresponding FDT.

Further, having anticipated, as in point (c) in Sec. \ref{sec2a2},
that an effective Eq. (\ref{e22}) does hold true, I demonstrate the
FDT directly for self-avoiding Rouse polymers in the following
manner. Imagine that I hold fixed the middle monomer of a
self-avoiding Rouse polymer, and keep taking snapshots of the polymer
configuration at fixed intervals of time, i.e., at $t=t_0, (t_0+\Delta
t), (t_0+2\Delta t),\ldots$. Afterwards, I take the snapshot at time
$t$ and evolve it $K$ number of times over a short time interval
$\delta t$ and note down the average displacement
$\displaystyle{\bar{\vec{\delta x}}}$ over these $K$ evolutions. A
look at Eq. (\ref{e22}) then tells us that this averaging process
kills the $\vec f_{N/2}$ term, and since $\vec v(t)\equiv0$ by
construction, the result is a quantity $\bar{\vec{\delta x}}$, which,
in the limit of $(K\rightarrow\infty,\delta t\rightarrow0)$, is simply
proportional to the average of $\vec g(t)$. These $\bar{\vec{\delta
    x}}$ values for $t=t_0, (t_0+\Delta t), (t_0+2\Delta t),\ldots$
can then be used to calculate $\mu(t-t')\equiv\langle\vec
g(t)\cdot\vec g(t')\rangle_{\vec v=0}$, proxied by
$\langle\bar{\vec{\delta x}}(t)\cdot\bar{\vec{\delta
    x}}(t')\rangle$. Such an exercise was indeed performed in
Ref. \cite{rousepaper}: using a Monte Carlo lattice polymer model with
$N=400$, $\Delta t=50$, $K=10^6$ and $\delta t=1$ (see
Ref. \cite{rousepaper} for model details). The corresponding result,
confirming the $t^{-2\nu/(1+2\nu)}$ power-law behavior of $\mu(t)$ is
shown in Fig. \ref{fig1}.
\begin{figure*}
\begin{center}
\begin{minipage}{0.48\linewidth}
\includegraphics[width=0.8\linewidth,angle=270]{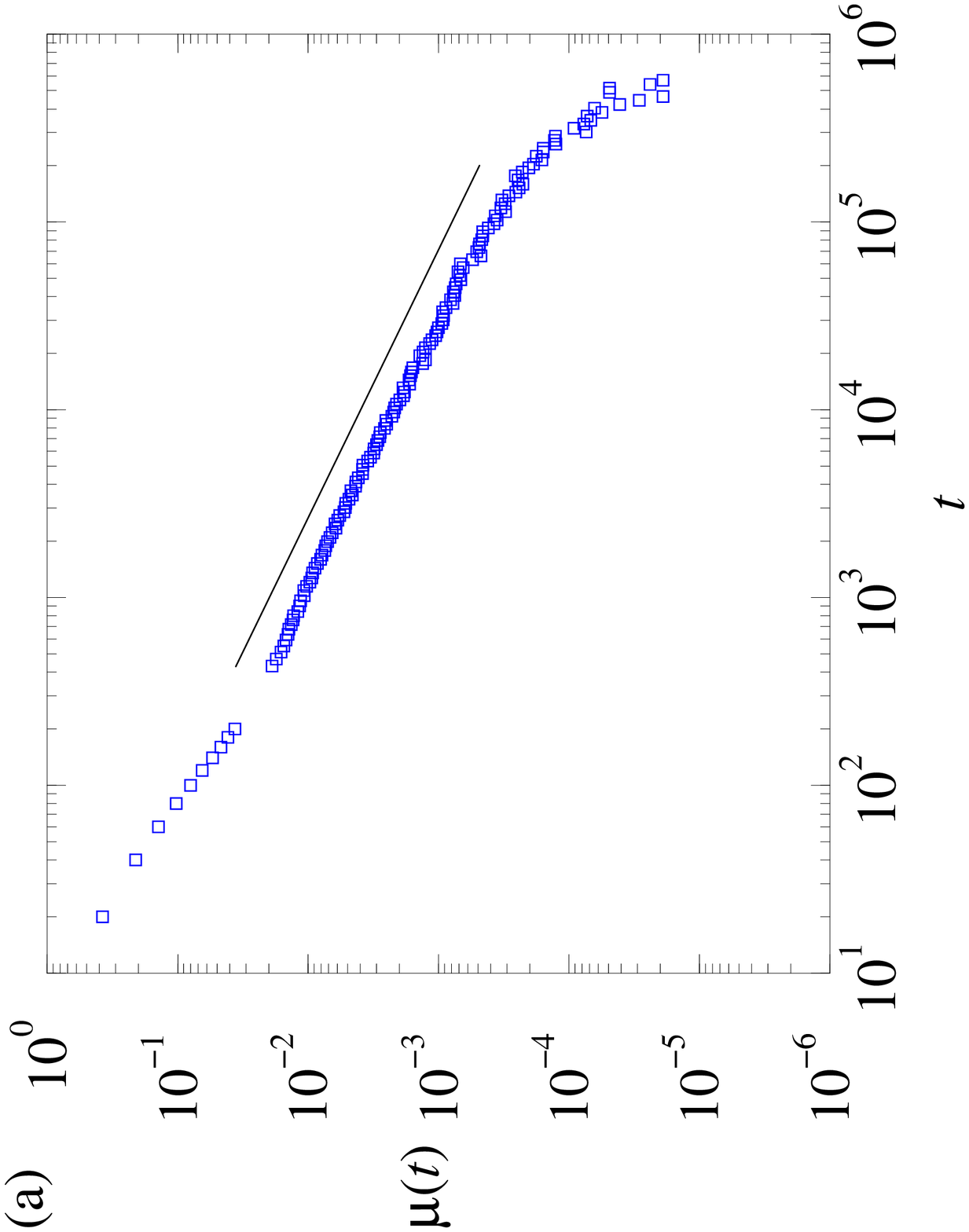}
\end{minipage} \hspace{5mm}
\begin{minipage}{0.48\linewidth}
\includegraphics[width=0.8\linewidth,angle=270]{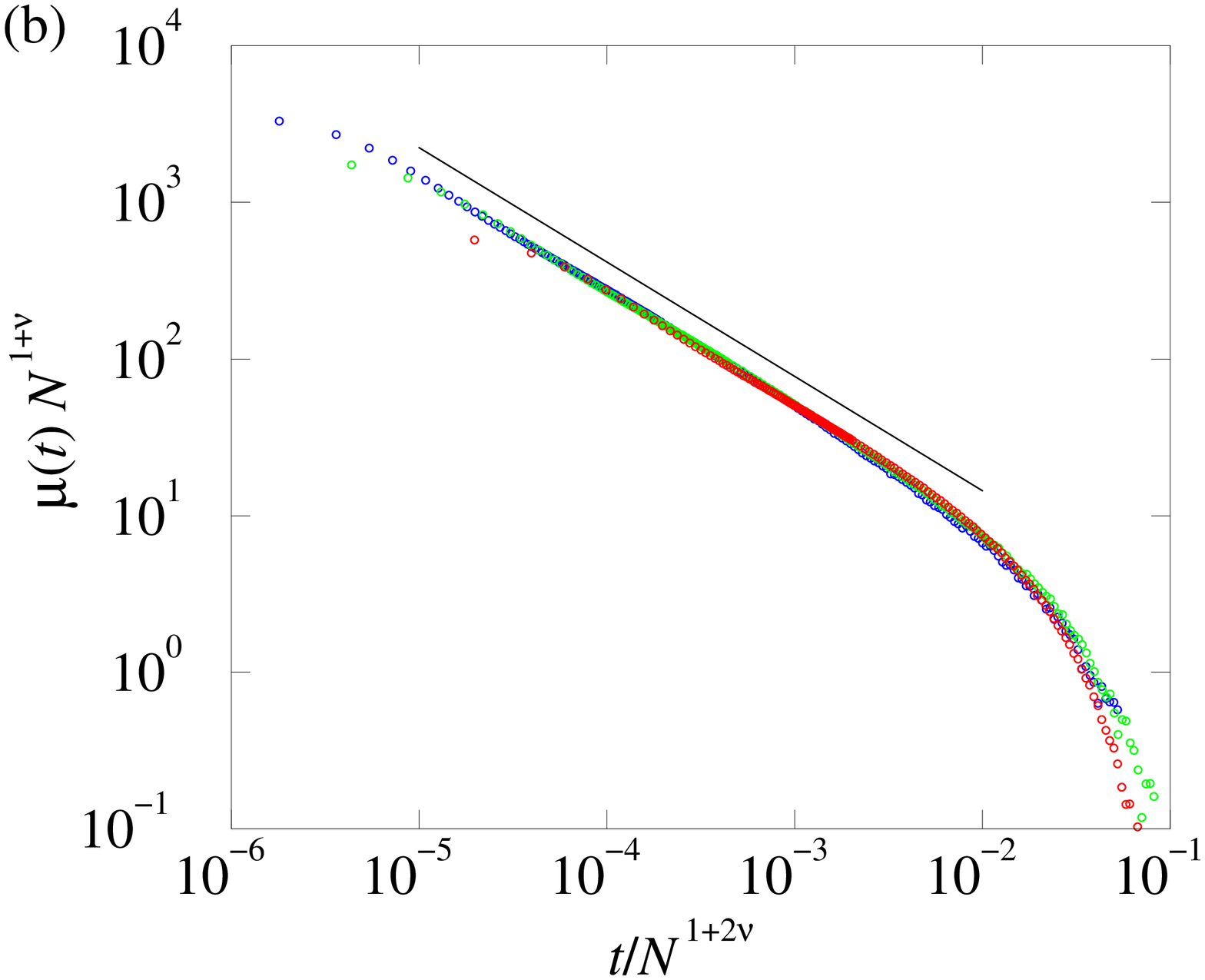}
\end{minipage}
\end{center}
\caption{(color online) The mean relaxation response $\sim
  t^{-(1+\nu)/(1+2\nu)}$ to local strain $\mu(t)$, created by
  injecting $n$ monomers at $t=0$ into a self-avoiding Rouse polymer
  tethered on a membrane, at the tether point. The data are obtained
  using a Monte Carlo lattice polymer model. (a) in 2D: for $N=250$,
  $n=5$. In 2D $(1+\nu)/(1+2\nu)=0.7$; the solid line corresponds to a
  power-law $t^{-0.7}$. (b) in 3D: for $n=10$ and $N=50$ (red), $100$
  (green) and $150$ (blue). In 3D $(1+\nu)/(1+2\nu)\approx0.73$; the
  solid line corresponds to a power-law $t^{-0.73}$. The choice of the
  scaling variable also confirms the terminal exponential decay of
  $\mu(t)$ behaving $\sim\exp(-t/\tau)$, with $\tau\sim N^{1+2\nu}$
  for self-avoiding Rouse polymers. The figures are taken from
  Refs. \cite{anom2} and \cite{anom1}, where the details of the
  polymer model and simulation details can be found, with permissions
  from Institute of Physics Publishing Ltd., UK (original articles
  published on 25 January 2008 and 8 October 2007
  respectively).\label{fig2}}
\end{figure*}

Given that if $\mu(t)\sim t^{-\alpha}$ then the anomalous dynamics
exponent is also $\alpha$, the well-known scaling of the MSD $\sim
t^{(2\nu)/(1+2\nu)}$ for self-avoiding Rouse and $\sim t^{2/3}$ for
self-avoiding Zimm polymers up to time $\tau$ and $\sim t$ thereafter
is trivially reproduced from the GLE formulation (\ref{e1}-\ref{e2})
\cite{rousepaper,rubbook} .

\subsection{The GLE formulation (\ref{e1}-\ref{e2}) for the anomalous
dynamics  of unbiased polymer translocation through a narrow pore in a
membrane\label{sec2c}}

Polymer translocation is a process where a polymer passes through a
narrow pore in a membrane. Of interest here is the so-called unbiased
(i.e., in the absence of any force or field) translocation: the
polymer passes through the pore purely due to thermal fluctuations,
and the dynamics is anomalous \cite{dprl}. For this system, below I
now characterize $\mu(t)$ as the polymers' mean local strain
relaxation response, as described in point (a) of Sec. \ref{sec2a2},
in the following manner.

A translocating polymer consists of two sub-polymers --- one on each
side of the membrane --- exchanging monomers through the pore. When a
monomer translocates, the polymer locally stretches on the side the
monomer translocates from, and locally compresses on the
other. Consequently, the polymer's chain tensions at the pore changes:
it increases on the side of the membrane which the monomer
translocates from, and decreases on the other. The relevant mean
polymeric response therefore, is to a (local) strain due to extra
monomers injection at the tether point of a polymer tethered on a
membrane.

Consider the case where $n$ extra monomers are injected at $t=0$ into
a polymer tethered on a membrane, at the tether point. First, it has
been shown in Ref. \cite{dpre} that for phantom Rouse polymers the
mean response to such a strain is given by $\mu(t)\sim
t^{-1}e^{-t/\tau}$, with $\tau\sim N^2$. For self-avoiding polymers
$\mu(t)$ is obtained as follows. Given that the terminal time $\tau$
for a tethered polymer of length $N$ scale $\sim N^{1+2\nu}$ for a
Rouse and $\sim N^{3\nu}$ for a Zimm polymer, at time $t$, counting
away from the tether point, all the monomers within a backbone
distance $n_t\sim t^{1/(1+2\nu)}$ for a Rouse, and $\sim t^{1/(3\nu)}$
for a Zimm polymer, equilibrate to the injected monomers. The real
space extent of $n_t$ monomers is $r(n_{t})\sim n_t^\nu$, but since
the rest $(N-n_t)$ monomers are not equilibrated to the injected
monomers at time $t$, there are $(n_t+n)$ monomers squeezed in a space
that extends only to $r(n_{t})$. The corresponding compressive force
[force $=$ (spring constant) $\times$ (stretching distance)] from
these $(n_{t}+n)$ monomers, felt at the pore, and hence $\mu(t)$, is
the given by $\sim n_t^{-2\nu}[\delta r(n_{t})]\sim
n_t^{-2\nu}\,n[\partial r(n_{t})/\partial n_t]=\nu nn_{t}^{-(1+\nu)}$,
which scales $\sim t^{-(1+\nu)/(1+2\nu)}$ for a Rouse and $\sim
t^{-(1+\nu)/(3\nu)}$ for a Zimm polymer. (Once again, this behavior
lasts only till the terminal time $\tau$.) The mean relaxation
response to the local strain for a self-avoiding Rouse polymer in 2
and 3D using a Monte Carlo lattice polymer model --- originally
explicitly evaluated in Refs. \cite{anom1,anom2} --- are shown in
Fig. \ref{fig2} (see Refs. \cite{anom1,anom2} for the model details).

The above thus implies that the anomalous dynamics for polymer
translocation is described by the GLE, resulting in the scaling of the
MSD $\sim t^{(1+\nu)/(1+2\nu)}$ for self-avoiding Rouse and $\sim
t^{(1+\nu)/(3\nu)}$ for self-avoiding Zimm polymers up to time $\tau$
and $\sim t$ thereafter. Consequently, the pore-blockade time scales
$\sim N^2$ for a phantom Rouse \cite{muthu}, $\sim N^{2+\nu}$ for
self-avoiding Rouse \cite{anom1,anom2}, and $\sim N^{1+2\nu}$ for
self-avoiding Zimm \cite{anom1,kap} polymers.

\subsection{The GLE formulation (\ref{e1}-\ref{e2}) for a reptating
polymer\label{sec2d}}

The notion of polymer reptation was originally introduced by De Gennes
in the context of gel electrophoresis \cite{degennes71}. Gel
electrophoresis is a method to isolate DNA fragments. In a typical
experiment, agarose powder is dissolved in water, and upon waiting for
a sufficient time the agarose forms the cross-links of a gel. The sizes
of the pores within the gel is controlled by the concentration of
agarose. When DNA molecules are injected into this gel, they collect
their charge from the solution, and with the application of an
electric field --- typically of strengths of a few volts per cm ---
the DNA molecules are driven through the pores in the gel. For a given
pore size of the gel, the mobility of the DNA molecules caused by the
electric field crucially depends on their lengths, leading to
length-dependent  segregation, which in turn allows one to separate
the DNA strands by their lengths. The DNA persistent length under a
typical gel electrophoresis experimental condition is in the range
400-800 \AA, which is typically the size of the pores in the gel
itself. In other words, the gel prevents the polymer to move
transverse to its configuration, and only the longitudinal, or
curvilinear, motion of the DNA contributes to its motion under the
applied electric field --- this is the central idea of polymer
reptation.
\begin{figure}[h]
\includegraphics[width=0.9\linewidth]{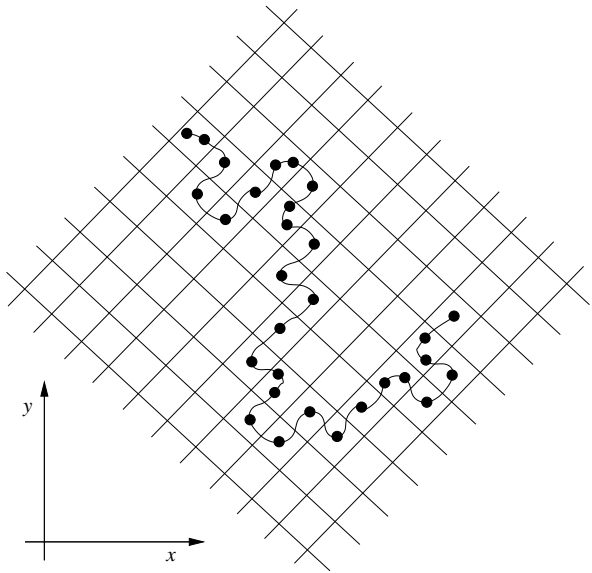}
\caption{A two-dimensional lattice representation of the repton
model. The pores are represented by cells formed by a square lattice,
while the polymer is represented by points (``reptons'') connected by
bonds. The rules governing this model are described in the
text.\label{fig3}}
\end{figure}

I consider here the three-dimensional repton model, one of the
simplest models with explicit reptation (longitudinal/curvilinear)
moves. The model was originally conceived by Rubinstein
\cite{rubinstein} to describe the motion of entangled polymers, and
was later co-opted by Duke \cite{duke} to describe the mobility of DNA
polymers in gel electrophoresis. A two-dimensional lattice
representation of this model is shown in Fig. \ref{fig3}: the network
of the agarose gel is represented by the crossing points of the grid
lines of the lattice, whereas the cells of the lattice correspond to
the gel pores. The DNA polymer, represented by points connected by
bonds, resides in the gel. The points are called ``reptons'': along
the backbone of the polymer two consecutive reptons are one persistent
length apart. This allows more than one repton to occupy the same
cell, but simultaneously two neighboring reptons cannot be more than
one cell apart. The movements of the reptons are given by the
following dynamical rules. A repton in the interior of the polymer can
move to one of its adjacent cells, provided one of its neighbor
reptons is already in that cell, and the other neighbor repton is in
the cell it leaves. The two end reptons can move into any of the
neighboring cells, so long as the bond between the end repton and its
neighbor is separated by not more than one cell.

The dynamical rules of the repton model ensures that while the
transverse motion for a repton in the interior of the polymer remains
blocked, it can only move towards (and away from) its neighbors: this
encodes explicitly the longitudinal or curvilinear motion of the
polymer in the repton model. Further, in any spatial dimension, the
curvilinear conformation of a reptating polymer can be described by a
string of 0s and 1s, corresponding to whether two consecutive monomers
reside within the same cell or not --- this forms the ``curvilinear
co-ordinate'' description of the polymer. (E.g., the curvilinear
co-ordinate description of the two-dimensional polymer in
Fig. \ref{fig3}, from the left end to the right, is given by the
sequence $01111101101111110111111011101$). The 0s in the curvilinear
co-ordinate description of the repton model is also known as the
``stored lengths'', as they correspond to local compression of the
polymer: the number of stored lengths within any given lattice site is
clearly one less than the number of monomers occupying that site. The
dynamics of the repton model is then summarized by the two following
simple rules: within the polymer motion takes place only by exchanging
a 0 with a 1, while at the ends of the polymer, a zero can become a 1,
or vice versa. The equilibrium density of stored lengths is a
parameter in the repton model, for the simulation results reported in
this paper, this parameter was chosen to be $1/3$.

\begin{figure}[h]
\includegraphics[width=0.9\linewidth]{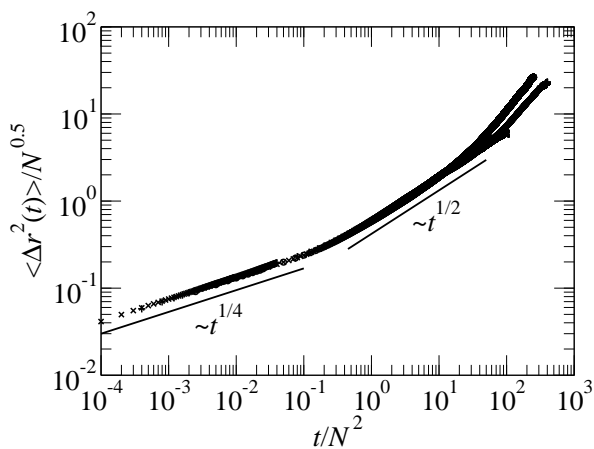}
\caption{The MSD $\langle\Delta r^2(t)\rangle$ of the middle repton
  for polymers with $N$ reptons: $N=201$ (circles), $501$ (pluses) and
  $1001$ (crosses). Data averaged over 1,024 independent polymer
  realizations. Upto time $\approx0.25N^2$ the MSD behaves as
  $t^{1/4}$, after which it crosses over to $\sim t^{1/2}$
  behavior. The onset of the diffusive behavior at time $\sim N^3$ is
  not shown, although increased slopes at late times for shorter
  polymers can be seen clearly.\label{fig4}}
\end{figure}

The dynamics of reptation is anomalous \cite{de}; i.e., for a polymer
of $N$ reptons the MSD of the middle repton increases $\sim t^{1/4}$
up to time $\sim N^2$; thereafter, the MSD for the middle monomer
increases $\sim t^{1/2}$ till time $\sim N^3$, after which the motion
becomes diffusive. I confirm this anomalous dynamical behavior of the
middle repton in Fig. \ref{fig4}. By the time the MSD behavior becomes
diffusive, the middle monomer displaces itself by the size of the
polymer, and since by construction the repton model describes a
phantom polymer (i.e., in equilibrium the size of a reptating polymer
of $N$ reptons scales $\sim N^{1/2}$), the diffusion coefficient of
the polymer scales as $1/N^2$ \cite{prahofer,drew}. In fact, given
that the polymer's motion is only curvilinear, and the polymer's
contour itself performs a random walk in space, one realizes that if
this anomalous dynamics of the middle monomer is translated back into
curvilinear co-ordinates, then the MSD of the middle repton increases
$\sim t^{1/2}$ up to time $\sim N^2$, and becomes diffusive
thereafter. It is this anomalous dynamics of the middle monomer in
curvilinear co-ordinates that I show to stem from the GLE
(\ref{e1}-\ref{e2}). In particular, I demonstrate that the forces on
the middle monomer in the curvilinear co-ordinate, at zero middle
repton velocity, satisfy the FDT, leading one to $\mu(t)\sim
t^{-1/2}\exp(-t/\tau)$ with $\tau\sim N^2$. As described in point (b)
of Sec. \ref{sec2a2}, this establishes the GLE formulation
(\ref{e1}-\ref{e2}) for the repton model.

\vspace{4mm}
\begin{figure}[h]
\includegraphics[width=0.9\linewidth]{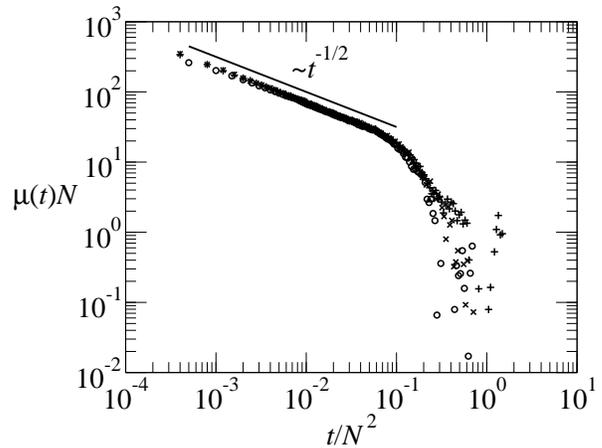}
\caption{The $\sim t^{-1/2}\exp(-t/\tau)$ behavior of
$\mu(t)$, obtained via the FDT in the curvilinear co-ordinate for the
anomalous dynamics of the middle repton for three different values of
$N$: $N=201$ (circles), $501$ (pluses) and $1001$ (crosses). Data averaged
over 1,024 independent polymer realizations. The solid line
corresponds to the power-law $t^{-1/2}$. The choice of the scaling
variable also confirms the terminal exponential decay of $\mu(t)$
behaving $\sim\exp(-t/\tau)$, with $\tau\sim N^2$. \label{fig5}}
\end{figure}

The above implies that the anomalous dynamics for polymer
reptation is described by the GLE, resulting in the scaling of the MSD
in the curvilinear co-ordinate $\sim t^{1/2}$ till time $\tau$ and
diffusive thereafter, as confirmed in Fig. \ref{fig4}.  In the
curvilinear co-ordinate, the polymeric motion is described by the
motion of the stored lengths or the 1s, which undergo the so-called
tagged-particle diffusion \cite{drew,bark}. At zero velocity of the
middle repton, the force it experiences is proportional to the
curvilinear gradient of stored lengths at its location. In order to
obtain the correlation in the forces on the middle repton at zero
velocity, I number the reptons $1,2,\ldots,N$, hold the middle repton
(tagged by $N/2$) fixed, and monitor the difference
$g(t)=\rho_{N/2,5}(t)-\rho_{N/2,-5}(t)$, where $\rho_{m,n}(t)$ is the
of number of stored lengths between reptons $m$ and $m+n$. The
correlation in the force on the middle repton at zero velocity is then
proxied by $\langle g(t)g(t')\rangle$. The expected behavior of
$\mu(t-t')\propto \langle g(t)g(t')\rangle\sim t^{-1/2}\exp(-t/\tau)$
is shown in Fig. \ref{fig5}, with $\tau\sim N^2$.
 
\vspace{4mm}

\section{The GLE formulation (\ref{e1}-\ref{e2}) for the anomalous
dynamics of polymer melts\label{sec3}}

\subsection{Polymer melts and reptation theory\label{sec3a}}

Mobility of individual polymers in systems containing long polymers
decreases with increasing density, causing their dynamics to slow
down. A notable manifestation of slow dynamics in dense polymeric
systems such as polymer melts is that their viscosity scales with
their molecular weight as a power law, with the viscosity exponent
$m_\eta$; experimentally measured to be $3.4\pm 0.2$ over an
impressive range of molecular weights and chemical compositions
\cite{lodge}. The fact that the slow dynamics of dense polymeric
systems can actually be attributed to the polymers becoming entangled,
especially if the polymers are long, was the seminal idea of De Gennes
\cite{degennes}. He argued that entanglement restricts the polymers to
only sliding past each other --- thereby reducing their mobility ---
as it is impossible for them to slide across each other. Indeed, the
dynamics of a melt can be quantitatively understood by taking the
entanglement idea to analyze the motion of a single polymer in a melt
\cite{degennes,de} --- referred to as the tagged polymer henceforth.

To approach the scaling of $m_\eta$ by obtaining that of $D$, De
Gennes considered the melt to only represent a \emph{rigid static}
network for the tagged polymer, with the tagged polymer reptating
through this network. Such a formulation reduces the many-polymer
problem of a melt to an {\it effective single (tagged) polymer
problem}, wherein the movements of the tagged polymer is restricted
only to reptation, i.e., stored length transportation due to
(longitudinal) fluctuations along its contour. By using the longest
available time-scale $\tau_r\sim N^2$ for the stored length
distribution within the tagged polymer's contour in this reduced
problem, he obtained $D\sim N^{-2}$, i.e., $m_\eta=3$
\cite{degennes,de,degennes71}. [The discrepancy between the
predictions of reptation theory and the experimental value for
$m_\eta$ remains to this day, and is beyond the scope of this paper.]
\begin{figure}[h]
\begin{center}
\includegraphics[width=0.9\linewidth]{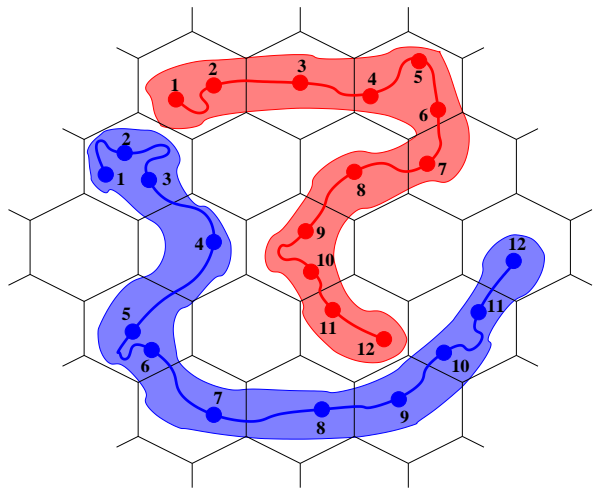}
\end{center}
\caption{(color online) Illustration of the two-dimensional version of
the lattice polymer model. Polymers are shown by darker colors and
their contours in lighter colors. In the upper polymer, interior
monomers 2, 4, 6, 9, 10 and 11 can either move along the contour, or
move sideways; monomer 7 can join either 6 or 8; the end monomers 1
and 12 can move to any empty nearest-neighbor site.  In the lower
polymer, interior monomers 3, 5, 6, 10 and 11 can either move along
the contour, or move sideways; monomer 1 can move to any empty
nearest-neighbor site, and monomer 12 can join its neighbor 11. In
this configuration, because of the self- and mutually-avoiding
property of the polymers, all other monomers cannot make a
move. Statistically once per unit of time, each monomer attempts to
move up and down along the contour, as well as sideways.
\label{fig6}}
\end{figure}

The prediction of reptation theory for the anomalous dynamics of the
middle monomer of the tagged polymer in a melt, beyond th so-called
entanglement time $\tau_e$ is the same as that for the repton model
discussed in Sec. \ref{sec2d}: at times $\gg\tau_e$, the MSD of the
middle monomer increases $\sim t^{1/4}$ up to time $\sim N^2$;
thereafter, it increases $\sim t^{1/2}$ till time $\sim N^3$, after
which the motion becomes diffusive. Such a picture is consistent with
the fact that by the time the dynamics becomes diffusive, the middle
monomer displaces itself by the size of the polymer, which scales as
$\sim N^{1/2}$ in a melt \cite{degennes}, yielding the result $D\sim
N^{-2}$. For the GLE description of the anomalous dynamics for
a many-body problem like a polymer melt, I make use of computer
simulations to demonstrate the FDT for the force experienced by the
middle monomer of the tagged polymer, as described below.

\subsection{Simulation details\label{sec3b}}

The simulations have been performed with the lattice polymer model
described in Ref. \cite{heuk}. This model combines a very high
computational efficiency with realistic polymer dynamics. All the
simulations are three-dimensional, with the lattice polymers residing
on a face-centered-cubic lattice. Polymer contours are self- and
mutually-avoiding. Monomers adjacent in the string are located either
in the same, or in neighboring lattice sites. Multiple occupation of
lattice sites is not allowed, except for a string of adjacent monomers
belonging to the same polymer. A two-dimensional version of the model
is illustrated in Fig. \ref{fig6}.

\vspace{4mm}
\begin{figure}[h]
\begin{center}
\includegraphics[width=0.9\linewidth]{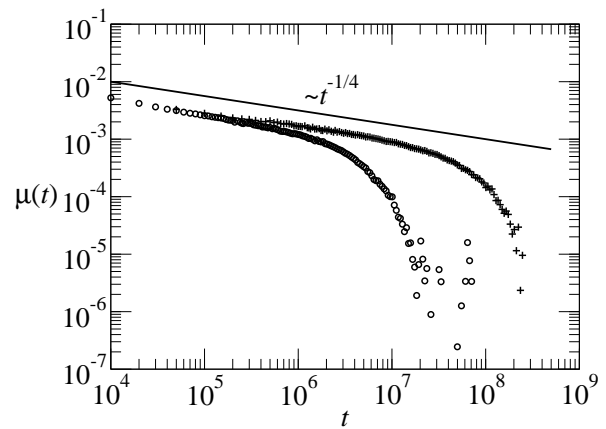}
\end{center}
\caption{$\mu(t)$ for the middle monomer of a tagged polymer in a
  polymer melt: $N=750$ (circles), $1500$ (pluses). Data averaged over
  512 independent polymer melt realizations. The solid line
  corresponds to the power-law $t^{-1/4}$. Beyond the entanglement
  time $\tau_e$, the data are consistent with an initial power-law
  behavior $\mu(t)\sim t^{-1/4}$. \label{fig7}}
\end{figure} 
\begin{table*}
\begin{tabular}{c|c|c}
  \hline\hline $\quad$polymeric system$\quad$ & $\quad\mu(t)\quad$&
  $\quad$$\langle \vec v(t)\rangle$$\quad$ \cr \hline\hline
  self-avoiding Rouse  & $\sim t^{-2\nu/(1+2\nu)}\exp(-t/\tau)$& $\sim
  t^{-1/(1+2\nu)}\vec F$ till $\tau$ and $\sim\vec F$ thereafter \cr
  &&$\langle[\vec x(t)-\vec x(0)]\rangle\sim
  t^{2\nu/(1+2\nu)}\vec F$
  till $\tau$, and $\sim t\vec F$ thereafter \cr\hline
  self-avoiding Zimm  & $\sim t^{-2/3}\exp(-t/\tau)$ & $\sim
  t^{-1/3}\vec F$ till $\tau$ and $\sim\vec F$ thereafter  \cr &&
  $\langle[\vec x(t)-\vec x(0)]\rangle\sim t^{2/3}\vec F$ till $\tau$ and
  $\sim t\vec F$ thereafter (not shown
  here)\cr\hline  reptation; repton model & $\sim
  t^{-1/2}\exp(-t/\tau)$ &$\sim t^{-1/2}$ till $\tau$ and $\sim t$
  thereafter \cr (curvilinear co-ordinate)&& $\langle[x(t)-
  x(0)]\rangle\sim  t^{1/2} F$ till $\tau$ and $t\sim F$
  thereafter\cr\hline melt (reptation theory) &$\sim t^{-1/4}$ between
  $\tau_e$ and $\sim N^2$&$\sim t^{-3/4}$ between $\tau_e$ and $\sim
  N^2$ (*) \cr & (as much as data can resolve)& $\langle[\vec
  x(t)-\vec x(0)]\rangle$ $\sim t^{1/4}\vec F$ between $\tau_e$ and $\sim
  N^2$ (not shown here) \cr\hline\hline
\end{tabular}

\vspace{4mm}
\begin{tabular}{c|c|c}
  \hline\hline $\quad$polymeric system$\quad$ & $\quad\mu(t)\quad$&
  $\quad$ number of translocated monomers in time $t$$\quad$ \cr
  \hline\hline phantom Rouse & $\sim t^{-1}\exp(-t/\tau)$ & $\sim t$
  \cr field-driven polymer translocation (3D) &($\tau$ not
  translocation time) & translocation time $\sim N$ (not shown here)
  \cr\hline self-avoiding Rouse  & $\sim
  t^{-(1+\nu)/(1+2\nu)}\exp(-t/\tau)$ & $\sim t^{(1+\nu)/(1+2\nu)}$
  \cr field-driven polymer translocation (3D) & ($\tau$ not
  translocation time) & translocation time $\sim N^{(1+2\nu)/(1+\nu)}$
  \cr\hline self-avoiding Zimm & $\sim
  t^{-(1+\nu)/(3\nu)}\exp(-t/\tau)$ & $\sim t^{(1+\nu)/(3\nu)}$ \cr
  field-driven polymer translocation & ($\tau$ not translocation
  time)& translocation time $\sim N^{3\nu/(1+\nu)}$ (not shown
  here)\cr\hline self-avoiding Rouse  & $\sim t^{-1/2}\exp(-t/\tau)$ &
  $\sim t^{1/2}$ \cr pulled polymer translocation & $\tau\sim N^2$
  ($\tau$ not translocation time) & translocation time $\sim N^2$
  \cr\hline \hline
\end{tabular}

\vspace{4mm}
\begin{tabular}{c|c|c}
  \hline\hline $\quad$polymeric system$\quad$ & $\quad\mu(t)\quad$&
  $\quad$ number of monomers adsorbed in time $t$$\quad$ \cr
  \hline\hline phantom Rouse & $\sim t^{-1}\exp(-t/\tau)$ & $\sim t$
  \cr polymer adsorption (3D) & ($\tau$ not adsorption time) &
  adsorption time $\sim N$ (not shown here) \cr\hline self-avoiding
  Rouse  & $\sim t^{-(1+\nu)/(1+2\nu)}\exp(-t/\tau)$ & $\sim
  t^{(1+\nu)/(1+2\nu)}$ \cr polymer adsorption (3D) & ($\tau$ not
  adsorption time)  & adsorption time $\sim N^{(1+2\nu)/(1+\nu)}$
  \cr\hline self-avoiding Zimm & $\sim
  t^{-(1+\nu)/(3\nu)}\exp(-t/\tau)$ & $\sim t^{(1+\nu)/(3\nu)}$ \cr
  polymer adsorption & ($\tau$ not adsorption time) & adsorption time
  $\sim N^{3\nu/(1+\nu)}$ (not shown here)\cr\hline \hline
\end{tabular}
\caption{The expected drift results, from the analysis in
  Secs. \ref{sec4a} and \ref{sec4b} --- when the systems start evolving
  at $t=0$ under the application of weak forces --- following the results
  of Table \ref{table1}. The top table relates to Sec. \ref{sec4a} and
  the bottom two tables relate to Sec. \ref{sec4b}. Note in Table
  \ref{table2} that if $\mu(t)\sim
  t^{-\alpha}$, then the drift exponent is $\alpha$. Note also that
  not all results are explicitly demonstrated in Secs. \ref{sec4a} and \ref{sec4b} using
  computer simulations; these are clearly marked. (*) $\tau_e$ is the
  entanglement time for polymer melts within the reptation
  theory \cite{de}.\label{table2}}
\end{table*}

The simulations are performed for a system of size $60^3$ with an
overall monomer density unity per lattice site. The polymers move
through a sequence of random single-monomer hops to neighboring
lattice sites. These hops can be along the contour of the polymer,
thus explicitly providing reptation dynamics. They can also change the
contour ``sideways'', providing Rouse dynamics (Fig. \ref{fig6}). Each
monomer attempts to move along the contour, as well as sideways
stretching or reducing the backbone, \emph {statistically once per
unit of time}. Due to the possibility that adjacent monomers belonging
to the same polymer can occupy the same site, overall approximately
40\% of the sites typically remain empty. The number of stored lengths
within any given lattice site is one less than the number of monomers
occupying that site.

Initial thermalizations were performed as follows: completely crumpled
up polymers are placed in lattice sites at random. The system is then
brought to equilibrium by letting it evolve for $\tau_{\rm  eq}(N)$
units of time, with a combination of intermediate redistribution of
stored lengths within a polymer \cite{redist}; for the polymer lengths
concerned here $\tau_{\rm eq}(N)>10^9$. The entanglement time for this
model is $\approx10^5$.

\subsection{The FDT for the forces on the middle monomer of the tagged
polymer in a polymer melt\label{sec3c}}

First, I anticipate, as in point (c) in Sec. \ref{sec2a2}, that an
effective Eq. (\ref{e22}) does hold true for the middle monomer of the
tagged polymer in a polymer melt. Then I follow the same procedure
along the as I did for demonstrating the FDT for the case of
self-avoiding Rouse polymers; namely, I hold the middle monomer fixed
and keep taking snapshots of the polymer configuration at fixed
intervals of time, i.e., at $t=t_0, (t_0+\Delta t), (t_0+2\Delta
t),\ldots$. Afterwards, I take the snapshot at time $t$ and evolve it
$K$ number of times over a short time interval $\delta t$ and note
down the average displacement $\bar{\vec{\delta x}}$ over these $K$
evolutions. Once again, a look at Eq. (\ref{e22}) tells us that this
averaging process kills the $\vec f_{N/2}$ term, and since $\vec
v(t)\equiv0$ by construction, the result is a quantity
$\bar{\vec{\delta x}}$, which, in the limit of
$(K\rightarrow\infty,\delta t\rightarrow0)$, is proportional to the
average of $\vec g(t)$. These $\bar{\vec{\delta x}}$ values for
$t=t_0, (t_0+\Delta t), (t_0+2\Delta t),\ldots$ can then be used to
calculate $\mu(t-t')\equiv\langle\vec g(t)\cdot\vec g(t')\rangle_{\vec
  v=0}$, proxied by $\langle\bar{\vec{\delta
    x}}(t)\cdot\bar{\vec{\delta x}}(t')\rangle$.

The corresponding result for $N=750$ and $1500$ for
$\Delta t=10,000$ and $50,000$ respectively, with $K=100,000$ and
$\delta t=1$ is shown in Fig. \ref{fig7}. Beyond the entanglement time
$\tau_e$, the data are consistent with an initial power-law behavior
$\mu(t)\sim t^{-1/4}$ (but are not sufficient to conclude what follows
afterwards). Nevertheless, this exercise provides a clear indication
that the anomalous dynamics of the middle monomer of the tagged
polymer is described by the GLE between the  entanglement time
$\tau_e$ and $\sim N^2$.

\section{Drifts in polymeric systems driven by weak forces: analogue
of the Nernst-Einstein relation\label{sec4}}

Until now, I have considered the classical polymeric systems with
anomalous dynamics in the absence of external driving forces, and
demonstrated that their anomalous dynamics are described by a unified
GLE scheme (\ref{e1}-\ref{e2}) with power-law memory kernels. I now
demonstrate that their GLE formulation is robust: it also adequately
describes drifts in polymeric systems driven by weak forces, as it
should.  

\vspace{6mm}
\begin{figure}[h]
\begin{center}
\includegraphics[width=0.9\linewidth]{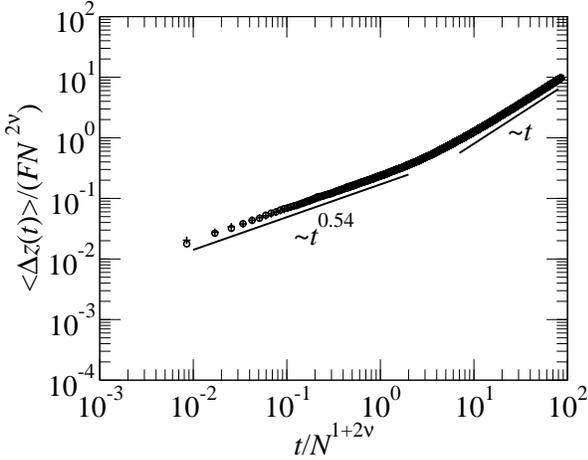}
\end{center}
\caption{Confirmation of Eq. (\ref{e31}) for an end monomer of a
  self-avoiding Rouse polymer: data collapse for the displacement
  $\langle\Delta z(t)\rangle$ in the $\hat z$-direction, under the
  application of a force $\vec F=F\hat z$ from $t=0$ onwards. Data are
  obtained using the same Monte Carlo lattice polymer model that was
  used to generate Fig. \ref{fig1}: $N=300,
  F\lambda/(k_BT)=0.075$ (circles) and $N=400, F\lambda/(k_BT)=0.1$
  (pluses), where $\lambda$ is the lattice constant. Data averaged
  over 8,192 equilibrated polymer realizations at $t=0$. Following Eq.
  (\ref{e31})$, \langle\Delta z(t)\rangle$ is expected to behave as
  $t^{2\nu/(1+2\nu)}F\approx t^{0.54}F$ till the terminal relaxation
  time $\tau\sim N^{1+2\nu}$, after which it should behave as
  $tF$. Note the labeling of the axes: they confirm that in the steady
  state ($t>\tau$) the drift scales as $1/N$ as it should.
  \label{fig12}}
\end{figure} 
The expected drift results --- when the polymeric system starts to
evolve at $t=0$ under the application of external forces --- are shown
in Table \ref{table2}. Note in Table \ref{table2} that if $\mu(t)\sim
t^{-\alpha}$, then the drift exponent is also $\alpha$: I elaborate on
this below. Note also that not all results in Table \ref{table2} are
demonstrated in Secs. \ref{sec4a} and \ref{sec4b} using computer
simulations; these are clearly marked.

\subsection{The standard extension of the GLE formulation to driven
polymeric systems\label{sec4a}}

The external constant force $\vec F$ is switched on a specified
monomer over an equilibrated polymer ensemble at $t=0$, and is assumed
to be weak enough that it does not distort the memory kernel [or the
FDT for $\vec g(t)$]. With these assumptions --- if the system's
dynamics is not rate-limited by some other process ---
Eqs. (\ref{e1}-\ref{e2}) retain their forms, however, Eq. (\ref{e4a})
needs to be modified to include the external force, so that it becomes
\begin{eqnarray}        
\gamma\vec v(t)=\vec\phi(t)+\vec F+\vec f(t),
\label{e28}
\end{eqnarray}      from where I have dropped the subscript $N/2$ from
$\vec f(t)$.  Together with Eqs. (\ref{e1}-\ref{e2}), Eq. (\ref{e28})
provides us with
\begin{eqnarray}        
\vec v(t)=\int_0^t dt' \,\beta(t-t')\,[\vec
g(t')+\vec F+\vec f(t')],
\label{e29}
\end{eqnarray}  
where $\beta(t)$ is defined in Eq. (\ref{e23}).

\vspace{6mm}
\begin{figure}[h]
\begin{center}
\includegraphics[width=0.9\linewidth]{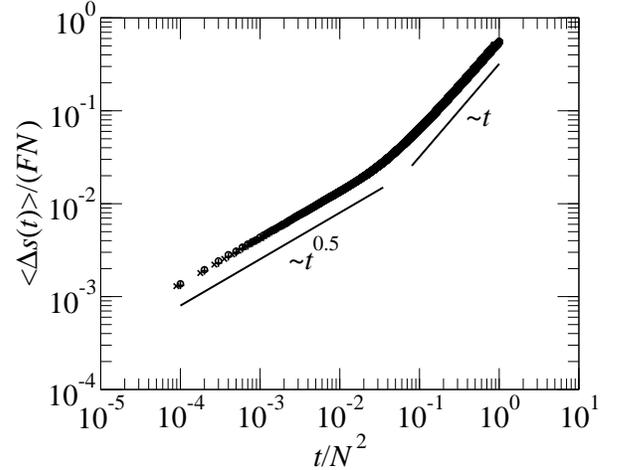}
\end{center}
\caption{Confirmation of Eq. (\ref{e31}) for the middle repton of the
  repton model. Data collapse for the curvilinear displacement
  $\langle\Delta s(t)\rangle$, under the application of a curvilinear
  force $F$ from $t=0$ onwards: $N=201, F\lambda/(k_BT)=0.1$
  (circles), $N=501, F\lambda/(k_BT)=0.25$ (pluses) and $N=1001,
  F\lambda/(k_BT)=0.5$ (crosses), where $\lambda$ is the lattice
  constant. Data averaged over 16,384 independent polymer realizations
  at equilibrium at $t=0$. Following Eq. (\ref{e31}) $\langle\Delta
  s(t)\rangle$ is expected to behave as $t^{1/2}F$ till the terminal
  relaxation time $\tau\sim N^2$, after which it should behave as
  $tF$. Note the labeling of the axes: they confirm that in the steady
  state ($t>\tau$) the drift scales as $1/N$ as it
  should. \label{fig13}}
\end{figure}
An ensemble average of Eq. (\ref{e29}) implies that
\begin{eqnarray}        
  \langle\vec v(t)\rangle=\int_0^t dt'
  \,\beta(t-t')\,\vec F.
\label{e30}
\end{eqnarray}  
With $\mu(t)\sim t^{-\alpha}\exp(-t/\tau)$ for some $\alpha$, at long
times (times still $\ll\tau$), Eq. (\ref{e30}) yields
\begin{eqnarray}        
  \langle\vec v(t)\rangle\sim t^{\alpha-1}\vec
  F; \quad\mbox{i.e.,}\quad \langle[\vec x(t)-\vec x(0)]\rangle\sim\vec
  F t^\alpha,
\label{e31}
\end{eqnarray}   
which lasts till time $\tau$, after which $\langle\vec x(t)\rangle$
increases as $t\vec F$. Such behavior can be thought of as the
analogue of the Nernst-Einstein relation \cite{vankampen}.

I now demonstrate Eq. (\ref{e31}) for a self-avoiding Rouse polymer,
and the repton model in the curvilinear co-ordinate.

\subsubsection{A self-avoiding Rouse polymer under a weak force on an
end monomer\label{se4a1}}

Using the Monte Carlo lattice polymer model that was used to generate
Fig. \ref{fig1} (details of the model can be found in
Ref. \cite{rousepaper}), a force $\vec F\equiv F\hat z$ was applied at
$t=0$ onwards on one of the end monomers of a set of equilibrated
self-avoiding Rouse polymer realizations at $t=0$, and the average
displacement $\langle \Delta z(t)\rangle$ of this monomer was tracked
as a function of time. The result, presented in Fig. \ref{fig12},
confirms Eq. (\ref{e31}). Note the labeling of the axes in
Fig. \ref{fig12}: they confirm that in the steady state ($t>\tau$) the
drift scales as $1/N$ as it should.

\subsubsection{The repton model with a weak curvilinear force on the
middle repton\label{sec4a2}}

Similarly, for the repton model a curvilinear force $F$ was applied
from $t=0$ onwards on the middle repton of a set of equilibrated
polymer realizations to $t=0$, and the average curvilinear
displacement $\langle\Delta s(t)\rangle$ of this monomer was tracked
as a function of time. The result, presented in Fig. \ref{fig13},
again confirms Eq. (\ref{e31}). Note the labeling of the axes in
Fig. \ref{fig13}: they confirm that in the steady state ($t>\tau$) the
drift scales as $1/N$ as it should.

\subsection{A different extension of the GLE description to driven
polymeric systems\label{sec4b}}

If the external force is not acting on a specified monomer (or the
system's dynamics is rate-limited by some other process), then a
different generalization of the GLE (\ref{e1}-\ref{e2}) is required to
obtain the drifts. This holds for dynamics of field-driven
translocation in 3D \cite{fieldtrans}, polymer adsorption on a solid
surface in 3D \cite{adsorb}, and polymer translocation by means of a
pulling force $F$ \cite{pulledtrans}. For the first two cases, the
force does not act on a specific monomer; instead, it acts
respectively on the monomer that is in the pore, and on the monomer
that is just in contact with the adsorbing surface. However, for
polymer translocation by means of a pulling force, the force does act
on an end monomer, but the translocation dynamics is rate-limited by
polymer dynamics at the pore. For these problems, a description based
on Eq. (\ref{e1}) alone is sufficient, with $\vec\phi(0)\neq0$,
provided, once again, that the force $\vec F$ is small enough that it
does not distort the memory kernel or the FDT for $\vec g(t)$. In that
case, Eq. (\ref{e1}) reads
\begin{eqnarray}   \vec\phi(t)=\vec\phi(0)-\int_0^tdt'\mu(t-t')\vec
v(t')+\vec g(t),
\label{e32}
\end{eqnarray}  leading one to
\begin{eqnarray}        \langle\vec v(t)\rangle=\int_0^t dt'
\,a(t-t')\,\langle[\vec\phi(0)-\vec\phi(t')]\rangle,
\label{e33}
\end{eqnarray}  
which is essentially an extension of
Eq. (\ref{e2}). If $\langle\vec\phi(t)\rangle$ becomes a constant
fairly soon after the force starts acting on the polymer, then from
Eq. (\ref{e33}) one has
\begin{eqnarray}   \langle\vec x(t)-\vec x(0)\rangle\sim t^\alpha
\label{e34}
\end{eqnarray} upto time $\tau$, and $\sim t$ thereafter. 

\vspace{6mm}
\begin{figure}[h]
\begin{center}
\includegraphics[angle=270,width=0.9\linewidth]{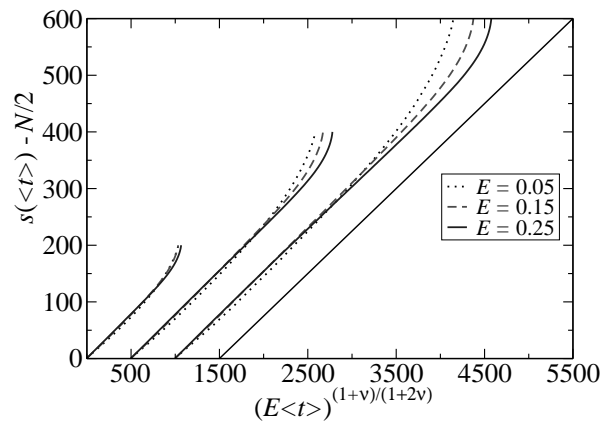}
\end{center}
\caption{Demonstration of the time scaling of Eq. (\ref{e34}) for
  polymer translocation driven by a field $E$ in 3D. Reproduced from
  Ref. \cite{fieldtrans} with permission from Institute of Physics
  Publishing Ltd., UK (original article published on 15 February
  2008). A note about the choice of variables: in order to avoid
  saturation effects \cite{fieldtrans} the average time $\langle
  t\rangle$, required to have $s$ monomers translocated for the first
  time, is plotted (instead of plotting the average number of
  translocated monomers $\langle s(t)\rangle$ as a function of $t$)
  for $N=400$ (average over $16,000$ polymer realizations for each
  field), $N=800$ (average over $16,000$ polymer realizations for each
  field), and $N=1,200$ ($5,000$ polymer realizations for
  $E\lambda/(k_BT)=0.05$, and $7,500$ polymer realizations each for
  $E\lambda/(k_BT)=0.15$ and $E\lambda/(k_BT)=0.25$). Here $\lambda$
  is the lattice constant. The data for $N=800$ correspond to real
  time value, while the data for $N=400$ and $N=1,200$ have been
  shifted by $\mp 500$ units along the x-axis for clarity. The solid
  line has been added for a guide to the eye. \label{fig8}}
\end{figure}
I now demonstrate the time scaling of Eq. (\ref{e34}) for
field-driven translocation of a Rouse polymer, adsorption of a Rouse
polymer on a solid surface, and translocation of a Rouse polymer by
means of a pulling force.

\subsubsection{Field-driven polymer translocation and polymer
adsorption on a solid surface\label{sec4b1}}

With the memory kernel $\mu(t)$ for a translocating polymer shown in
Fig. \ref{fig2}, I proceed directly to the simulation data to
demonstrate Eq. (\ref{e34}) for field-driven translocation 
(translocation driven by a field $E$ that only acts on the monomer in
the pore) and polymer adsorption on a solid surface (with energy
of adsorption $\varepsilon$ per monomer), both in 3D. These results
were originally reported, using a Monte Carlo lattice polymer code, in
Refs. \cite{fieldtrans} and \cite{adsorb}, where the reader can find
the details on the simulations. The force felt by the monomer in the
pore for translocation, and by the monomer in contact with the
adsorbing surface, respectively, are directed perpendicular to the
membrane and the adsorbing surface. Consequently, their dynamics is
described simply by using the (scalar) components of the velocities
and forces on the monomers perpendicular to the membrane and the
adsorbing surface.
\begin{figure}[h]
\begin{center}
\includegraphics[width=0.7\linewidth]{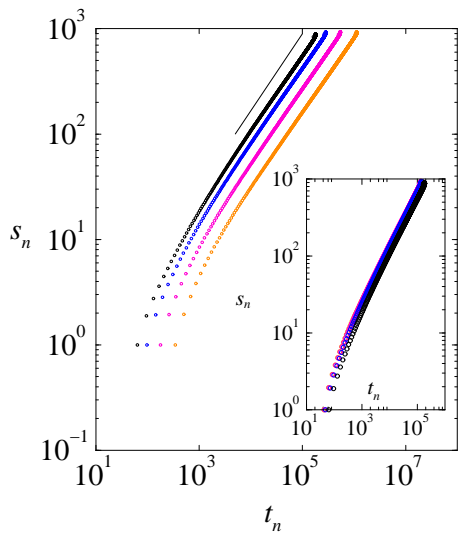}
\end{center}
\caption{(color online) Demonstration of the time scaling of
  Eq. (\ref{e34}) for polymer adsorption on a surface. Figure
  reproduced from Ref. \cite{adsorb} with permission from Institute of
  Physics Publishing Ltd., UK (original article published on 12 May
  2009). A note about the choice of variables: in order to avoid
  saturation effects \cite{adsorb}, (instead of plotting the average
  number of adsorbed monomers as a function of time) the average
  number of adsorbed monomers $s_n$ in time $t_n$ is plotted, where
  $t_n$ is the time have $n$-th monomer adsorbed for the first
  time. Adsorption data with weak adsorption energies for $N=1000$
  (from left to right) $\varepsilon=2$ (black), $\varepsilon=3$
  (blue), $\varepsilon=4$ (magenta) and $\varepsilon=5$ (orange): the
  data are progressively separated by a factor $2$ along the $x$-axis
  for clarity. The original data are shown in the inset in the same
  color scheme. The solid black line corresponds to an exponent
  $(1+\nu)/(1+2\nu)\simeq0.73$. \label{fig9}}
\end{figure}

For field-driven translocation in 3D, a polymer of length $N$  was
threaded fixed halfway through the pore and was thermalized with the
field $E$ switched on, acting only on the monomer in the pore
perpendicular to the membrane (this makes $\phi(0)\neq0$). Then
translocation was started at $t=0$. It was shown in
Ref. \cite{fieldtrans} that $\langle\vec\phi(t)\rangle$ approaches a
constant fairly quickly.

Similarly, for polymer adsorption on a solid surface in 3D, a polymer
of length $N$ was thermalized with one end held tethered on the
surface. For this case, entropy-related stretching of a polymer close
to a surface makes $\phi(0)\neq0$. Adsorption was started at $t=0$
(see Ref. \cite{adsorb} for further details).

The data for field-driven polymer translocation and polymer adsorption
in 3D, are reproduced from Refs. \cite{fieldtrans} and \cite{adsorb}
in Figs. \ref{fig8} and \ref{fig9} respectively; both confirm
Eq. (\ref{e34}). It however needs to be mentioned here that for
field-driven polymer translocation and polymer adsorption to a solid
surface in 2D, the memory kernel is overruled by conservation of
energy: for these problems, the number of monomers translocated and
the number of monomers adsorbed respectively scale $\sim
t^{1/(2\nu)}$, i.e., the translocation and the adsorption time scale
as $N^{2\nu}$. The reader can find more details on this in
Ref. \cite{anom2}.

\subsubsection{Polymer translocation by a pulling force\label{sec4b2}}

The setup for polymer translocation by a pulling force is as
follows. A polymer is threaded through a narrow pore in a membrane,
and a latex bead is attached to one of the end monomers. An optical
tweezer captures the bead, and pulls the polymer perpendicularly away
from the membrane, facilitating translocation of the polymer through
the pore \cite{pullingexp}. Here I consider translocation of a
self-avoiding Rouse polymer, with the membrane placed on the
$yz$-plane, while the bead is pulled with a constant force $F$ along
the $+x$ direction, causing the polymer to translocate from left to
right. As mentioned earlier, the translocation is rate-limited by the
events at the pore, so the action of the force $F$ alone does not
determine the translocation dynamics.

It is clear from this setup that the polymer on the right of the
membrane would be stretched due to the action of the force, but not on
the left of the membrane. Therefore, although for this setup the
power-law part of the memory kernel for the polymer on the left of the
membrane is still given by $\sim t^{-(1+\nu)/(1+2\nu)}$, one needs to
establish the memory kernel for the stretched part of the polymer on
the right of the membrane. Indeed, as I show below, the power-law part
of the memory kernel for the stretched part of the polymer on the
right of the membrane is given by $t^{-1/2}$: it is once again
obtained from the mean relaxation of a (local) strain in the
polymer at the pore, when the strain is caused by monomer insertion
into a stretched polymer tethered on a membrane at the tether
point. In order to do so, one does need two ingredients: (i) the
terminal decay time $\tau$ of a stretched polymer of length $N$
tethered on a membrane scales $\sim N^2$ \cite{pulledtrans}, (ii) the
shape of stretched polymer is that of a cylinder, where the radius of
the cylinder is given by that of the Pincus blob $\xi$. The magnitude
of the pulling force $F$ determines the value of $\xi$, in terms of
which the spring constant of a stretched polymer of length $N$ is
given by $\sim k_BT/(N\xi)$ \cite{degennes}. 

\begin{figure}[h]
\begin{center}
\includegraphics[angle=270,width=0.9\linewidth]{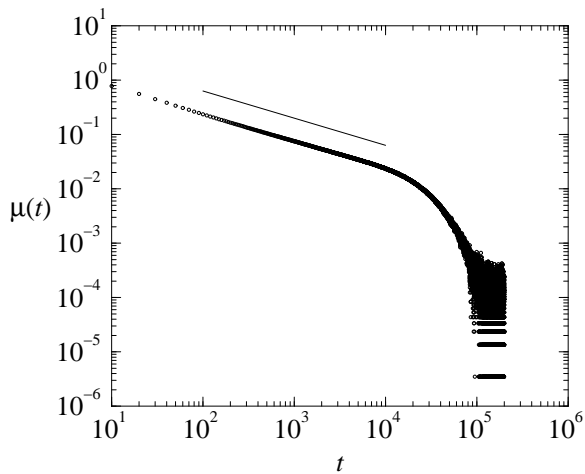}
\end{center}
\caption{Mean relaxation response of a stretched self-avoiding Rouse
  polymer, tethered on a membrane, to a local strain caused by monomer
  injection at the tether point at $t=0$. The data are obtained for
  $N=100$ and $F\lambda/(k_BT)=1$, using a Monte Carlo lattice polymer
  model. Here $\lambda$ is the lattice
  constant. The solid line corresponds to the power-law $\sim
  t^{-1/2}$. Figure reproduced from Ref. \cite{pulledtrans} with
  permission from Elsevier Inc.\label{fig10}}
\end{figure}

Consider the case when $n$ extra monomers are injected at $t=0$, at
the tether point of a self-avoiding Rouse polymer, which is tethered
on a membrane at one end, and is stretched by a force applied at the
open end. Given (i-ii) above, at time $t$, counting away from the
tether point, all the monomers within a backbone distance $n_t\sim
t^{1/2}$ will equilibrate to the new situation. However, since the
polymer is stretched, its shape is that of a cylinder, and the real
space extent of $n_t$ monomers is $r(n_{t})\sim n_t$, but since the
rest $(N-n_t)$ monomers are not equilibrated to the injected monomers
at time $t$, there are $(n_t+n)$ monomers squeezed in a space that
extends only to $r(n_{t})$. The corresponding compressive force [force
$=$ (spring constant) $\times$ (stretching distance)] from these
$(n_{t}+n)$ monomers, felt at the tether point, and hence $\mu(t)$, is
the given by $\sim [\delta r(n_{t})]/(n_t\xi)\sim n[\partial
r(n_{t})/\partial n_t]/(n_t\xi)=n/(n_t\xi)$, which scales $\sim
t^{-1/2}$ \cite{pulledtrans}. (Once again, this behavior lasts only
till the terminal time $\tau$.) The explicitly evaluated $t^{-1/2}$
power-law behavior of the mean relaxation response of a stretched
self-avoiding Rouse polymer, tethered on a membrane, to a local strain
caused by monomer injection at the tether point is shown in
Fig. \ref{fig10}. The data in Fig. \ref{fig10}, obtained by using a Monte
Carlo lattice polymer model, are reproduced from
Ref. \cite{pulledtrans}, from where the reader can find the details of
the model.

\begin{figure}[h]
\begin{center}
\includegraphics[angle=270,width=0.9\linewidth]{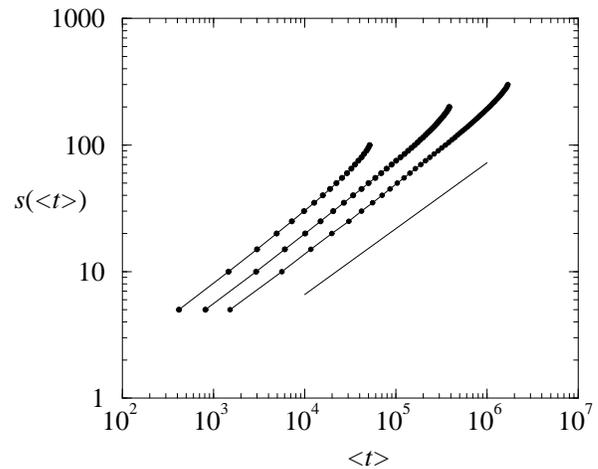}
\end{center}
\caption{Mean time required, for $F\lambda/(k_BT)=1$, to translocate
  $s$ monomers for the first time, for $s=5,10,15,\ldots,N$: from left
  to right $N=200$, $N=400$, $N=600$. The time-axis corresponding to
  $N=400$ is the true time, for the $N=200$ and $N=600$ cases the time
  axis is divided and multiplied by a factor $2$ respectively. The
  angular brackets denote an average over $48,000$ polymer
  realizations. A note about the choice of variables: in order to
  avoid saturation effects \cite{pulledtrans} the average time
  $\langle t\rangle$ required to have $s$ monomers translocated for
  the first time is plotted (instead of plotting the average number of
  translocated monomers $\langle s(t)\rangle$ as a function of
  $t$). The solid line corresponds to an exponent $1/2$. Figure
  reproduced from Ref. \cite{pulledtrans} with permission from
  Elsevier Inc. \label{fig11}}
\end{figure}

Having put all the above together, one finds that for translocation of
a self-avoiding Rouse polymer with a pulling force, there are two
different (power-law) memory kernels for the polymer on the two sides
of the membrane: on the left it is $t^{-(1+\nu)/(1+2\nu)}$ and on the
right it is $t^{-1/2}$. The system's dynamics is determined by the
slower of the two: i.e., by $t^{-1/2}$. Thus, when the polymer is
threaded fixed halfway through the pore and thermalized with the force
on one of the end monomers, $\phi(0)\neq0$, and one expects
Eq. (\ref{e34}) to hold true, with $\alpha=1/2$. This is verified,
using a Monte Carlo lattice polymer model in Fig. \ref{fig11}. The
data are reproduced from Ref. \cite{pulledtrans}.

\section{A critique on the use of fractional Fokker-Planck equation to
describe anomalous dynamics for polymer translocation\label{sec5}}

In this short section, I return to the issue of non-Markovian property
of the anomalous dynamics in polymeric systems.

The fact that in polymeric systems, the motion of a tagged monomer is
non-Markovian is as such not surprising: in 3D, a polymer with $N$
monomers resides in a phase space of $6N$ dimensions (in
many-polymeric systems such as polymer melts, the dimension of the
phase space is 6 times the entire number of monomers in the systems),
while the dimension of the phase space associated with a tagged
monomer is only 6. Thus, if one is to follow the dynamics of a tagged
monomer, one is essentially taking a projection of the entire system's
dynamics on to a few degrees of freedom. While the dynamics of the
entire system is no doubt Markovian, a projection on to a few selected
degrees of freedom can indeed render the dynamics non-Markovian
--- this is the very foundation of the GLE \cite{mori,mazo}.

As I have shown in this paper, the dynamics of a tagged monomer in a
wide range of single-polymeric (phantom Rouse, self-avoiding Rouse,
self-avoiding Zimm, reptation, translocation through a narrow pore in
a membrane), as well as many-polymeric (polymer melts) systems is
robustly formulated by the GLE (\ref{e1}-\ref{e2}). It is also worth
emphasizing that --- as apparent from Eq. (\ref{e22}) --- in this GLE
formulation if the concerned monomer makes a move at any time, there
is an enhanced chance to undo this move in subsequent times; this is
where anomalous dynamics in polymeric systems stem from. Anyhow, the
GLE describes trajectories in the phase space. Given this, a question
that naturally arises is: ``how does one formulate a probabilistic
description of the trajectories in phase space for anomalous polymer
dynamics?'' What I am appealing to is that the Fokker-Planck equation
is a probabilistic formulation of the trajectories described by the
Langevin equation; so what would be an appropriate probabilistic
description of the trajectories described by Eqs. (\ref{e1}-\ref{e2}),
or for that matter Eq. (\ref{e22})?

While for now I will leave this question for future research here,
this section will not be complete without mentioning that in a limited
context --- to describe the anomalous dynamics of polymer
translocation --- fractional Fokker-Planck equation (fFPE) has been
postulated \cite{fFPtrans}. In this equation, an extension of the
standard Fokker-Planck equation, anomalous dynamics is a consequence
of introducing power-law waiting times before each jump of the
concerned particle, as the jump length and the waiting times for any
jump is obtained from fixed probability distributions, independently
of their values at previous jumps. Given that there is no power-law
waiting time for the movements of the concerned monomer (instead if it
makes a move at any time, there is an enhanced chance to undo this
move in subsequent times; this is where anomalous dynamics in
polymeric systems stem from), not only that the application of fFPE to
describe polymer translocation is not correct, but also it remains a
far cry for describing the examples of anomalous dynamics in polymeric
systems that are considered in this paper.

\section{Conclusion\label{sec6}}

Extending the work of a recent Letter \cite{panja1}, in this
pedagogical paper I extensively elaborated on the fact that the
anomalous dynamics of a tagged monomer in a wide range of
single-polymeric (phantom Rouse, self-avoiding Rouse, self-avoiding
Zimm, reptation, translocation through a narrow pore in a membrane),
as well as many-polymeric (polymer melts) systems is robustly
formulated by the GLE. In this GLE formulation the velocity $\vec
v(t)$ of a tagged monomer and the force $\vec\phi(t)$ it experiences,
are related to each other via
\begin{eqnarray}  
\vec\phi(t)=-\int_0^tdt'\mu(t-t')\vec v(t')+\vec
g(t).
\label{e35}
\end{eqnarray}   
In Eq. (\ref{e35}) $\mu(t)$ is the memory kernel, and the stochastic
noise term $\vec g(t)$ satisfies the condition that $\langle\vec
g(t)\rangle_0=0$, while the fluctuation-dissipation theorem (FDT)
$\langle\vec g(t)\cdot\vec
g(t')\rangle_0\equiv\langle\vec\phi(t)\cdot\vec\phi(t')\rangle_{\vec
  v=0}=3k_BT\mu(t-t')$ in 3D. Here $k_B$ is the Boltzmann constant,
$T$ is the temperature, and $\langle\ldots\rangle_0$ denotes an
average over the stochastic noise realizations, including an average
over equilibrium configurations of the polymers at $t=0$. Equation
(\ref{e35}) can be inverted to write
\begin{eqnarray}  
\vec v(t)=-\int_0^tdt'a(t-t')\vec\phi(t')+\vec h(t),
\label{e36}
\end{eqnarray}   
with $\tilde\mu(s)\tilde a(s)=1$ in the Laplace space, $\langle\vec
h(t)\rangle_0=0$, and the corresponding FDT $\langle\vec h(t)\cdot\vec
h(t')\rangle_0\equiv\langle\vec v(t)\cdot\vec v(t')\rangle|_{\vec
  \phi=0}=3k_BT\,a(t-t')$. On the one hand $\mu(t)$ is the mean
relaxation response of the polymers to local strains, and can be
derived from the equilibrium statistical physics of polymers; and on
the other, $a(t)$ characterizes the anomalous dynamics via the FDT: as
the mean-square displacement of a tagged monomer is obtained by
integrating $\langle\vec v(t)\cdot\vec v(t')\rangle_{\vec\phi=0}$
twice in time. An important property of the anomalous dynamics that
transpires through this exercise is that if $\mu(t)\sim t^{-\alpha}$
for some $\alpha$, then the anomalous dynamics is also $\alpha$. In
other words, the anomalous dynamics for polymeric systems are
connected to the mean relaxation response of the polymers to local
strains. The anomalous dynamics and the mean relaxation response of
the polymers to local strains, as shown in Eq. (\ref{e22}) works in
the following way: if the concerned monomer makes a move at any time,
there is an enhanced chance to undo this move in subsequent times;
this is where anomalous dynamics in polymeric systems stem from.

Further, the characteristics of the drifts caused by a (weak)
applied field on polymeric systems, too, are obtained from the
corresponding memory kernels: if $\mu(t)\sim t^{-\alpha}$ for some
$\alpha$, then the drift exponent is also $\alpha$. This could be
thought of as the analog of the Nernst-Einstein relation.

Given that the GLE provides the trajectory description in the phase
space, I bring to light the non-Markovian character of the anomalous
dynamics for polymeric systems. The fact that the motion of a tagged
monomer is non-Markovian is as such not surprising: in 3D, a polymer
with $N$ monomers resides in a phase space that has $6N$ dimensions
(in many-polymeric systems such as polymer melts, the dimension of the
phase space is 6 times the entire number of monomers in the systems),
while the dimension of the phase space associated with a tagged
monomer is only 6. Thus, if one is to follow the dynamics of a tagged
monomer, one is essentially taking a projection of the entire system's
dynamics on to a few degrees of freedom. While the dynamics of the
entire system is no doubt Markovian, such a projection on to a few
degrees of freedom can indeed render the dynamics non-Markovian ---
this is the very foundation of the GLE \cite{mazo,mori}.  Given this,
a question that naturally arises is: ``how does one formulate a
probabilistic description of the trajectories in phase space for
anomalous polymer dynamics?'' While for now I leave this question for
future research, I note that anomalous dynamics in polymeric systems
cannot be captured by, e.g., the fractional Fokker-Planck equation
(fFPE) which has been recently postulated \cite{fFPtrans} to describe
the anomalous dynamics of polymer translocation through a narrow pore
in a membrane. In this equation, an extension of the standard
Fokker-Planck equation, anomalous dynamics is a consequence of
introducing power-law waiting times before each jump of the concerned
particle, as the jump length and the waiting times for any jump is
obtained from fixed probability distributions, independently of their
values at previous jumps. Given that there is no power-law waiting
time for the movements of the concerned monomer (instead if it makes a
move at any time, there is an enhanced chance to undo this move in
subsequent times; this is where anomalous dynamics in polymeric
systems stem from), not only that the application of fFPE to describe
polymer translocation is not correct, but also it remains a far cry
for describing the examples of anomalous dynamics in polymeric systems
that are considered in this paper.

\noindent {\bf Acknowledgements:} I thank Gerard T. Barkema for
stimulating discussions and for considerable amount of help with the
simulations. Ample computer time from the Dutch national supercomputer
cluster SARA is also gratefully acknowledged.

\begin{widetext}
\appendix
\section*{Appendix: Derivation of $\langle\vec v(t)\cdot\vec
v(t')\rangle$ for phantom polymers [i.e., Eq. (\ref{e24})]}
\setcounter{equation}{0}
\renewcommand{\theequation}{A\arabic{equation}}

I start with the velocity autocorrelation function as in
Eq. (\ref{e24})
\begin{eqnarray} V(t,t')=\langle\vec v(t)\cdot\vec
v(t')\rangle=\int_0^tdt_1\,\beta(t-t_1)\int_0^{t'}dt_2\,\beta(t'-t_2)[\underbrace{\langle\vec
g(t_1)\cdot\vec g(t_2)\rangle}_{G(t_1,t_2)}+\underbrace{\langle\vec
f_{N/2}(t_1)\cdot\vec f_{N/2}(t_2)\rangle}_{\Delta(t_1,t_2)}],
\label{ea1}
\end{eqnarray} I am interested in the behavior of $\langle\vec
v(t)\cdot\vec v(t')\rangle$ in the limit of large $(t,t')$ with finite
$(t-t')$. In this limit I expect $V(t,t')\rightarrow
V(t-t')$. Nevertheless, I take a dual Laplace transform of $V(t,t')$:
one with Laplace variable $s$ for $t$, and the other with Laplace
variable $s'$ for $t'$.
\begin{eqnarray} \tilde V(s,s')=\int_0^\infty dt\,
e^{-st}\int_0^\infty dt'\,
e^{-s't'}V(t,t')\nonumber\\&&\hspace{-5.65cm}=\int_0^\infty dt\,
e^{-st}\int_0^\infty dt'\,
e^{-s't'}\int_0^tdt_1\,\beta(t-t_1)\int_0^{t'}dt_2\,\beta(t'-t_2)\,[G(t_1,t_2)+\Delta(t_1,t_2)].
\label{ea2}
\end{eqnarray} Upon having interchanged integration variables, I
rewrite $\tilde V(s,s')$ as
\begin{eqnarray} \tilde V(s,s')=\int_0^\infty dt\,
e^{-st}\int_0^\infty dt'\,
e^{-s't'}\int_0^tdt_1\,\beta(t-t_1)\int_0^{t'}dt_2\,\beta(t'-t_2)\,[G(t_1,t_2)+\Delta(t_1,t_2)]
\nonumber\\&&\hspace{-12.95cm}=\int_0^\infty
dt_1\,e^{-st_1}\int_0^\infty
dt_2\,e^{-s't_2}\,[G(t_1,t_2)+\Delta(t_1,t_2)]\int_{t_1}^\infty
dt\,e^{-s(t-t_1)}\,\beta(t-t_1)\int_{t_2}^\infty
dt'\,e^{-s'(t-t_2)}\,\beta(t'-t_2)
\nonumber\\&&\hspace{-12.95cm}=\tilde\beta(s)\tilde\beta(s')
\int_0^\infty dt_1\,e^{-st_1}\int_0^\infty
dt_2\,e^{-s't_2}\,[G(t_1,t_2)+\Delta(t_1,t_2)].
\label{ea3}
\end{eqnarray} With $G(t_1,t_2)=3k_BT\mu(|t_1-t_2|)$ and
$\Delta(t_1,t_2)=6\gamma k_BT\delta(t_1-t_2)$, I choose to represent
$\Delta(t-t')$ as
\begin{eqnarray} \Delta(t-t')=6\gamma k_BT\left\{
\begin{array}{lll}\lim_{\Delta\rightarrow0}\frac1\Delta\quad\mbox{if}\,\,\,|t-t'|\le\Delta/2\\\\0\quad\quad\quad\quad\quad\mbox{otherwise.}
\end{array}\right.
\label{ea4}
\end{eqnarray} Thereafter, having recalled that
$\tilde\beta(s)=[\gamma+\tilde\mu(s)]^{-1}$, I obtain
\begin{eqnarray} \tilde V(s,s')=3k_BT\,
\tilde\beta(s)\tilde\beta(s')\,\left[\frac{\tilde\beta^{-1}(s)}{s'-s}+\frac{\tilde\beta^{-1}(s')}{s-s'}\right]=3k_BT\,\left[\frac{\tilde\beta(s')}{s'-s}+\frac{\tilde\beta(s)}{s-s'}\right],
\label{ea5}
\end{eqnarray} which implies, assuming $t>t'$, that
\begin{eqnarray} \langle\vec v(t)\cdot\vec
v(t')\rangle=3k_BT\,\beta(t-t').
\label{ea6}
\end{eqnarray}

\end{widetext}


\begin{thebibliography}{10}
\bibitem{laplace} Note that $\tilde\mu(s)\tilde a(s)=1$ yields $\tilde
a(s)\sim(s+\tau^{-1})^{1/2}$, the inverse Laplace transform of which,
strictly speaking, does not exist. This can be traced back to
Eq. (\ref{e15}), where I have ignored the minimum distance between the
consecutive monomers providing the ultraviolet cutoff for $p$ for the
summation. In reality, the $t^{-1/2}$ behavior does not extend all the
way to $t=0$, implying that the inverse Laplace transform of $\tilde
a(s)$ is indeed given by $t^{-3/2}$.
\bibitem{redist} This was done in conformation with detailed
  balance.
\end{thebibliography}
\end{document}